\documentclass[singlecolumn,10pt,aps,pra,superscriptaddress,floatfix,tightenlines,nofootinbib,nobibnotes]{revtex4-2}

\usepackage[colorlinks,linkcolor=blue,urlcolor=blue,citecolor=blue]{hyperref}
\usepackage{xurl}
\usepackage{booktabs}
\usepackage{comment}
\usepackage{xcolor}

\usepackage{dsfont}
\usepackage{physics}
\usepackage{nicefrac}
\usepackage{mathtools}
\usepackage{amsmath, amssymb, amsfonts, amsthm}
\usepackage{thmtools, thm-restate}

\usepackage{algpseudocodex}
\usepackage{algorithm}

\usepackage[T1]{fontenc}
\usepackage[utf8]{inputenc}
\usepackage[scaled=0.9]{DejaVuSansMono}
\usepackage{listings}

\usepackage{xcolor}
\definecolor{keywordcolor}{rgb}{0.7, 0.1, 0.1}
\definecolor{tacticcolor}{rgb}{0.0, 0.1, 0.6}
\definecolor{commentcolor}{rgb}{0.4, 0.4, 0.4}
\definecolor{symbolcolor}{rgb}{0.0, 0.1, 0.6}
\definecolor{sortcolor}{rgb}{0.1, 0.5, 0.1}
\definecolor{stringcolor}{rgb}{0.5, 0.3, 0.2}

\newif\ifshowasciileancode

\newtheorem{theorem}{Theorem}
\newtheorem*{theorem*}{Theorem}

\DeclareMathOperator{\supp}{supp}

\DeclareMathOperator{\id}{id}

\renewcommand{\Tr}{\operatorname{Tr}}

\allowdisplaybreaks

\begin{document}

\title{Lean-Quantum: Toward AI-Assisted Formalization of Quantum Information}

\author{Kazumi Kasaura}
\email{kazumi.kasaura@sinicx.com}
\affiliation{OMRON SINIC X Corporation}
\affiliation{RIKEN}
\author{Kei Tsukamoto}
\affiliation{The University of Tokyo}
\author{Kento Mori}
\email{mori.kento@molsis.co.jp}
\affiliation{MOLSIS Inc.}
\author{Risa Mizuno}
\affiliation{The University of Tokyo}
\author{Takahiro Namatame}
\affiliation{Kyoto University}
\author{Yuta Oriike}
\affiliation{CyberAgent, Inc.}
\author{Masaya Taniguchi}
\affiliation{RIKEN}
\author{Sho Sonoda}
\email{sho.sonoda@riken.jp}
\affiliation{RIKEN}
\affiliation{CyberAgent, Inc.}
\author{Hayata Yamasaki}
\email{hayata.yamasaki@gmail.com}
\affiliation{The University of Tokyo}

\begin{abstract}
Quantum information theory is built on entropic quantities; among them, the sandwiched R\'enyi relative entropy is a fundamental divergence with various applications, and its data processing inequality (DPI) under quantum channels is a cornerstone result. In this work, we present a Lean 4 library for quantum information, designed as a reusable formal infrastructure for theoretical analysis. As a central demonstration of the library, we formalize the DPI for the sandwiched R\'enyi relative entropy for positive semidefinite operators on finite-dimensional quantum systems. The library provides a basis-independent operator-theoretic framework for finite-dimensional quantum mechanics compatible with the standard mathematical library \textit{Mathlib}, including reusable interfaces for finite-dimensional systems, states, channels, tensor products, partial traces, Choi operators, Kraus representations, and Stinespring representations. It also builds infrastructure for noncommutative trace inequalities, including operator monotonicity and convexity via the real continuous functional calculus, block-operator positivity, Hilbert--Schmidt operator spaces, Jensen's operator inequality, generalized perspectives, operator power means, and Lieb--Ando trace inequalities. On top of this framework, we formalize entropy-specific ingredients for the DPI: variational formulas for the sandwiched quasi-entropy via Young and reverse-Young inequalities, tensor-product compatibility of real powers, and Haar measures on unitary groups. Together, these components yield a Lean formalization of the DPI, give strong subadditivity as a corollary, and provide the last missing component needed to complete the Lean formalization of the generalized quantum Stein's lemma. More broadly, the development provides machine-checkable foundations for future formalized and AI-assisted research in quantum information theory.

The Lean library is available as Ref.~\cite{lean-quantum}.
\end{abstract}

\maketitle

\tableofcontents

\section{Introduction}

\subsection{Background}

Quantum information theory provides a mathematical framework for characterizing the optimal performance and fundamental limitations of quantum information processing~\cite{Watrous2018}. Entropic quantities play a central role in this framework: they quantify distinguishability, uncertainty, correlations, channel capacities, error exponents, and security parameters. Their structural properties often determine whether an information-processing theorem has a physically meaningful interpretation. Among these properties, the monotonicity of quantum divergences under quantum channels is particularly fundamental. If a quantum channel represents a physically realizable processing of quantum systems, then a physically consistent measure of distinguishability should not increase under the action of the channel.

The sandwiched R\'enyi relative entropy~\cite{muller2013quantum,wilde2014strong,jakvsic95entropic} is a central quantum divergence satisfying this principle. For positive semidefinite operators $\rho$ and $\sigma$ and for $\alpha \in (0,1) \cup (1,\infty)$, the sandwiched R\'enyi relative entropy $D_{\alpha}(\rho\|\sigma)$ is defined by
\begin{align}
    D_{\alpha}(\rho\|\sigma)
    &\coloneqq
    \frac{1}{\alpha-1}
    \log\left[\frac{Q_{\alpha}(\rho\|\sigma)}{\Tr[\rho]}\right],&
    Q_{\alpha}(\rho\|\sigma)
    &\coloneqq
    \Tr\left[
        \left(
            \sigma^{\frac{1-\alpha}{2\alpha}}
            \rho
            \sigma^{\frac{1-\alpha}{2\alpha}}
        \right)^{\alpha}
    \right],
\end{align}
whenever $0<\alpha<1$, and whenever $\alpha>1$ and $\supp[\rho]\subseteq\supp[\sigma]$; otherwise the standard extended-real convention sets the divergence to $+\infty$ for $\alpha>1$. For normalized states, $\Tr[\rho]=1$. This quantity interpolates and extends several operationally meaningful divergences and has become a standard tool in quantum hypothesis testing, strong converse analysis, one-shot information theory, and the study of quantum channel capacities. Its data processing inequality (DPI) states that, for every completely positive trace-preserving map $\Phi$,
\begin{align}
    D_{\alpha}(\Phi[\rho]\|\Phi[\sigma])
    \le
    D_{\alpha}(\rho\|\sigma),
    \qquad
    \alpha \in [1/2,1)\cup(1,\infty).
    \label{eq:intro-dpi}
\end{align}
The inequality was shown for $\alpha\in(1,2]$ by Refs.~\cite{muller2013quantum,wilde2014strong} and then established in its full valid range $\alpha\ge 1/2$ by Frank and Lieb in Ref.~\cite{FrankLieb2013}, while Ref.~\cite{beigi2013sandwiched} also provided an alternative proof for $\alpha\in(1,\infty)$.
The proof in Ref.~\cite{FrankLieb2013} reduces the DPI to joint convexity or concavity of $Q_{\alpha}$. This convexity statement is proved from a variational formula and Lieb--Ando trace inequalities. The passage from the convexity or concavity of $Q_{\alpha}$ to channel monotonicity then uses Stinespring dilation, normalized Haar averaging over a unitary group, Jensen's inequality, tensor-product additivity, and monotonicity of the logarithm.
Thus, the DPI for the sandwiched R\'enyi relative entropy is not only a cornerstone theorem of quantum information theory, but also a representative example of the analytic and operator-theoretic complexity that modern quantum information theory involves.

The development of quantum information science has also increasingly relied on software infrastructure. Numerical and symbolic software tools, including \textit{Qiskit}~\cite{Qiskit2024}, \textit{Qulacs}~\cite{Uano2021}, and \textit{Stim}~\cite{Stim2021}, are widely used for designing, simulating, and testing quantum circuits, quantum algorithms, and quantum protocols. 
These tools have expanded the scale and reliability of computational exploration in quantum information science, but they do not, by themselves, verify the proofs of the theorems being used.
As quantum information theory becomes more intricate, and as AI systems begin to assist in conjecture generation, proof search, and formal manipulation, it becomes increasingly important to develop infrastructure that can verify logical correctness directly.

Interactive theorem provers provide such infrastructure. Lean 4~\cite{moura2021lean} is a proof assistant in which every theorem is checked by a small trusted kernel, and every logical dependency must ultimately be reduced to previously verified definitions and theorems. Its mathematical library, \textit{Mathlib}~\cite{mathlib2020}, already contains extensive formalized material in algebra, topology, measure theory, functional analysis, operator theory, $C^\ast$-algebras, and continuous functional calculus. A Lean formalization is therefore not only a machine-checked record of a single proof. It is also a reusable mathematical interface: a precise library of definitions, lemmas, and proof methods that can be searched, combined, and extended without ambiguity. 
Such libraries are particularly important for AI-assisted theoretical research. AI systems can generate plausible definitions, proof sketches, and code fragments, but plausibility alone does not establish logical correctness. A Lean library provides a machine-checkable reference point: just as a compiler checks that a program is well typed, Lean's kernel checks that each accepted proof term has the claimed theorem as its type, using only explicit definitions and previously verified results. It therefore makes it possible to distinguish heuristic reasoning suggested by AI from reasoning that has been certified as a formal proof.

\subsection{Main Results and Technical Contributions}

In this work, we present a Lean 4 library for the theory of quantum information. The codebase is a reusable software library~\cite{lean-quantum}, not a one-off encoding of a single theorem. As a key demonstration of the library, we formalize the DPI for the sandwiched R\'enyi relative entropy for positive semidefinite operators on finite-dimensional quantum systems. The library's technical contributions are formal results and interfaces that were not previously available in \textit{Mathlib}: a basis-independent quantum-information interface for finite-dimensional systems, an operator-theoretic hierarchy of noncommutative trace inequalities, and a formalization-friendly proof for the DPI. More concretely, the development contains the following contributions.
\begin{enumerate}
    \item Our first contribution is a basis-independent formulation of finite-dimensional quantum information theory in Lean, without reducing all statements to matrices in a fixed basis. Mathematically, $d\times d$ complex matrices and linear operators on a $d$-dimensional Hilbert space are equivalent, but they lead to very different interfaces in Lean. A matrix-based formulation forces dimensions and basis choices to appear explicitly throughout theorem statements, even when the underlying quantum-information argument only requires any finite-dimensional system without specifying $d$. Our formulation instead represents quantum systems as finite-dimensional complex Hilbert spaces, operators as linear endomorphisms, states as positive semidefinite trace-one operators, and channels as completely positive trace-preserving maps. This makes it possible to state results in the usual coordinate-free language of quantum information theory; moreover, with this formulation, we can directly reuse \textit{Mathlib}'s operator-algebraic infrastructure for continuous linear maps, ordered star structures, $C^\ast$-algebras, spectra, and the continuous functional calculus. Building on this interface, we formalize tensor-product systems, partial traces, vectorization, basis-dependent transposition, Choi operators, Kraus representations, Stinespring representations, and the standard finite-dimensional equivalences among complete positivity, positivity of ampliations, Choi positivity, Kraus representations, and Stinespring representations. This coordinate-free interface was missing from existing quantum-information-related Lean libraries~\cite{Meiberg2024,he2026codesigningquantumcodestransversal,ren2026merleanagenticframeworkautoformalization,ehatamm2026endtoendformalizationquantumerror,kol2026machineverifiedproofquantumoptimizationconjecture} and is a necessary foundation for formalizing analytic proofs in quantum information theory, including the DPI.
    \item Our second contribution is a reusable hierarchy of formalized tools for noncommutative trace inequalities. We formalize operator monotonicity, antitonicity, convexity, and concavity through the real continuous functional calculus; block-matrix and block-operator positivity; Hilbert--Schmidt operator spaces; Jensen's operator inequality~\cite{hansen1982jensen}; generalized perspective functions~\cite{effros2009matrix,ebadian2011perspectives}; operator power means~\cite{kubo1980means}; Lieb--Ando trace inequalities~\cite{Carlen2010,NIKOUFAR2013531}; trace Young and reverse-Young inequalities; tensor-product compatibility of real powers; normalized Haar measures on unitary groups; twirling identities; and Jensen inequalities for Bochner integrals. These formalized components are not specific to the DPI. They provide reusable theorem-prover infrastructure for quantum entropic inequalities, hypothesis-testing theorems, converse bounds, and other arguments based on noncommutative trace inequalities.
    \item Our third contribution is a technical contribution to quantum information theory itself: rather than merely translating the existing proof into Lean, we provide an alternative proof strategy whose intermediate statements are both mathematically modular and Lean-compatible. In particular, while Ref.~\cite{FrankLieb2013} proves the variational formulas for $Q_{\alpha}$ through an Euler--Lagrange optimization argument on the positive cone, we give and formalize an alternative proof of these formulas based on trace Young and reverse-Young inequalities. This replaces differentiability and first-order optimality arguments for noncommutative trace functionals by spectral decompositions, scalar inequalities, and explicit equality cases. We then combine these variational formulas with the Lieb--Ando trace inequalities, Stinespring dilation, Haar-unitary averaging, Jensen inequalities for Bochner integrals, tensor multiplicativity, and monotonicity of the logarithm to obtain the DPI. A further technical point is that the DPI is first proved on the cone of positive definite operators, where inverse powers and inverse square roots are ordinary objects of the real continuous functional calculus. This separates the core noncommutative analytic argument from support projections and extended-real conventions. The extension to positive semidefinite operators is then formalized separately by introducing an extended-real-valued non-negative version of the divergence and proving its monotonicity under CPTP maps. Isolating the statements for positive definite operators yields a tractable and reusable collection of statements in the Lean library, while the final theorem gives the standard support-sensitive DPI for positive semidefinite operators.
\end{enumerate}

These contributions are new also at the level of formalized mathematics: the resulting library supplies machine-checked versions of operator-algebraic and quantum-information arguments that previously existed only as natural-language proofs. The point is not merely to encode a known endpoint inequality, but to formalize a reusable chain of definitions, representation theorems, analytic inequalities, and entropy-specific arguments with explicit hypotheses and machine-checkable dependencies.

\subsection{Design Choices and Formalization Methodology}

The formalization is shaped by a small number of design choices that recur throughout the paper. A central design choice is to work with operators on finite-dimensional Hilbert spaces rather than with matrices in terms of a fixed basis. This is mathematically natural for quantum information theory and technically important in Lean. Many of the relevant structures already exist in \textit{Mathlib} for continuous linear maps, $C^\ast$-algebras, ordered star rings, spectra, and continuous functional calculus. By aligning the formalization with these standard abstractions, we avoid repeated dimension bookkeeping and make the resulting theorems reusable beyond a fixed matrix representation. This abstraction also made it possible to formulate some components, such as parts of the L\"oewner--Heinz theorem~\cite{lowner1934monotone,heinz1951beitrage,Carlen2010}, in more general ordered $C^\ast$-algebraic settings equipped with real continuous functional calculus.

We also represent many mathematical conditions as predicates on ambient objects rather than as bundled subtypes whenever this better matches existing \textit{Mathlib} interfaces. Positivity, positive definiteness, density-operator conditions, complete positivity, and trace-preserving conditions are treated as hypotheses on operators or linear maps. This keeps theorem statements close to ordinary mathematical usage, where one fixes operators $\rho$ and $\sigma$ and then assumes conditions such as $\rho\ge0$, $\sigma>0$, and $\Tr[\rho]=1$. It also reduces coercion overhead and makes it easier to apply library results stated for ambient objects satisfying appropriate hypotheses.

Another important choice is to first prove the DPI on the positive definite cone and then extend the result to positive semidefinite operators by a separate argument. Natural-language proofs of the DPI without Lean often use negative powers, inverse square roots, or variational optimizers while leaving support conditions implicit. Lean requires every inverse power to be justified by an invertibility or spectral-domain hypothesis. Working on the positive definite cone keeps these operations inside the ordinary continuous functional calculus and avoids mixing the main analytic proof with support projections or extended-real conventions.
The positive semidefinite case is therefore handled separately: after proving the core positive-definite DPI, the development introduces an extended-real-valued non-negative divergence and formalizes its monotonicity under CPTP maps, using the corresponding regularization and support-sensitive boundary conventions.

The proof route for trace inequalities was chosen with the same goal of producing reusable interfaces. Rather than formalizing Lieb's concavity theorem and Ando's convexity theorem directly, we formalize a simpler route in Ref.~\cite{NIKOUFAR2013531} in which the required Lieb--Ando trace inequalities are derived from generalized perspective functions~\cite{effros2009matrix,ebadian2011perspectives} and operator power means~\cite{kubo1980means}. This choice changes the structure of the formalization. The main proof obligations, including inverse-square-root bookkeeping, contraction weights, Jensen's operator inequality~\cite{hansen1982jensen}, and compatibility with the continuous functional calculus, are concentrated in reusable intermediate theorems. The Hilbert--Schmidt operator-space construction then converts the resulting operator-mean inequalities for left and right multiplication into the trace inequalities needed for the DPI. This organization follows the analytic dependencies of the proof while producing components that can be reused independently of the sandwiched R\'enyi relative entropy.

These choices also determine our methodology for AI-assisted formalization.
Our effective workflow was not to ask AI tools to discover the right Lean statements from scratch. Instead, we decomposed the mathematical proof into small statements with stable interfaces, inserted theorem skeletons to expose the intended dependency graph, and then used AI assistance for local proof search, lemma discovery, and code fragments compatible with the fixed interfaces. Statement design remained a human-guided part of the formalization. This is why the codebase~\cite{lean-quantum} is organized as many named definitions and lemmas rather than as a small number of large theorems: the decomposition makes the development more robust, reusable, and amenable to machine assistance.

\subsection{Related Work}

Formalization of quantum information theory has become increasingly important as the theory becomes more technically involved. A representative example is the generalized quantum Stein's lemma~\cite{Brand_o_2008,brandao2010reversible,brandao2010generalization,Brandao2015,hayashi2025generalizedquantumsteinslemma,10898013}, which is a fundamental result on asymptotic hypothesis testing of null hypotheses given by independent and identically distributed (IID) quantum states against composite alternative hypotheses specified by sets of non-IID quantum states. In 2010, the lemma was originally proposed and analyzed in Ref.~\cite{brandao2010generalization}. In 2022, a logical gap was identified in part of the argument in an early version of Ref.~\cite{fang2025towards}, whose proof strategy was inspired by the analysis of the generalized quantum Stein's lemma in Ref.~\cite{brandao2010generalization}. This observation led Ref.~\cite{Berta_2023} to point out a corresponding logical gap in Ref.~\cite{brandao2010generalization}. A stronger version of the generalized quantum Stein's lemma, under weaker assumptions, was proved in Ref.~\cite{hayashi2025generalizedquantumsteinslemma}, while Ref.~\cite{10898013} also independently proved the lemma under the original assumptions of Ref.~\cite{brandao2010generalization}. A formalization of these arguments would help assess the correctness of the resulting proofs by making their assumptions and logical dependencies machine-checkable.

Prior Lean work~\cite{Meiberg2024,LeanQuantumInfo}, which motivated the present work, aims to formalize the proof of the generalized quantum Stein's lemma in Ref.~\cite{hayashi2025generalizedquantumsteinslemma}. That development uses a matrix-based formulation and formalizes the argument presented in Ref.~\cite{hayashi2025generalizedquantumsteinslemma}. However, the part corresponding to the DPI for the sandwiched R\'enyi relative entropy, which is used in Ref.~\cite{hayashi2025generalizedquantumsteinslemma} as a known result, remained represented by unverified components~\cite{Meiberg2024}. In Lean, such components are represented by \texttt{sorry}, a placeholder that allows a statement to be accepted temporarily without a proof checked by the Lean kernel. Our formalization is complementary to this prior work: it develops a more abstract operator-theoretic infrastructure and supplies the analytic ingredients needed to fill the corresponding DPI-related gap. The present codebase~\cite{lean-quantum} is therefore intended not only as a proof of a single inequality, but also as a bridge between abstract \textit{Mathlib}-compatible operator theory and existing quantum-information formalization efforts.

Beyond \textit{Mathlib} for formalizing mathematics, there are also broader efforts to formalize physics in Lean, such as Physlib~\cite{physlib}, into which the prior quantum-information formalization work~\cite{Meiberg2024,LeanQuantumInfo} has been integrated. Within quantum information science, recent works have also developed Lean formalizations for topics such as quantum computation, quantum error correction, and quantum optimization~\cite{he2026codesigningquantumcodestransversal,ren2026merleanagenticframeworkautoformalization,ehatamm2026endtoendformalizationquantumerror,kol2026machineverifiedproofquantumoptimizationconjecture}. Related formalizations have also been developed in proof assistants other than Lean~\cite{Hietala_2021,Bordg_2020,peng2022formallycertifiedendtoendimplementation,zhou2022coqqfoundationalverificationquantum}. Our codebase is complementary to these efforts and has particular value as reusable infrastructure for the theory of quantum information. In particular, the library's coordinate-free formulation of finite-dimensional systems, states, channels, tensor products, partial traces, trace inequalities, and entropic formulas can serve as common infrastructure for future formalizations connecting quantum information theory with more general formalized mathematics and physics.

\subsection{Applications}
\label{sec:intro-applications}

As an immediate application, the DPI yields strong-subadditivity-type inequalities. Taking the quantum channel in Eq.~\eqref{eq:intro-dpi} to be a partial trace gives, for suitable tripartite states and for $\alpha\ge 1/2$, inequalities of the form
\begin{align}
    D_{\alpha}\left(
        \rho_{ABC}
        \middle\|
        \rho_{AB}\otimes \frac{I_C}{d_C}
    \right)
    \ge
    D_{\alpha}\left(
        \rho_{BC}
        \middle\|
        \rho_B\otimes \frac{I_C}{d_C}
    \right).
    \label{eq:intro-generalized-ssa}
\end{align}
In the limit $\alpha\to 1$, the sandwiched R\'enyi relative entropy reduces to the Umegaki relative entropy, and Eq.~\eqref{eq:intro-generalized-ssa} recovers the usual strong subadditivity of the von Neumann entropy,
\begin{align}
    S(\rho_{ABC}) + S(\rho_B)
    \le
    S(\rho_{AB}) + S(\rho_{BC}),
\end{align}
in its standard equivalent form. Thus, the present formalization provides the main formal route toward one of the most fundamental structural inequalities in quantum information theory,
with the positive semidefinite setting handled through the extended-real-valued non-negative formulation and the corresponding continuity and support-regularization argument.

A second application concerns the formalization of the generalized quantum Stein's lemma. Existing Lean developments toward formalizing this theorem contained a substantial unverified component corresponding to the DPI of the sandwiched R\'enyi relative entropy~\cite{Meiberg2024,LeanQuantumInfo}. Since the DPI in our library is formalized for general finite-dimensional Hilbert spaces and operators, it is more general than a matrix-based formulation in a fixed basis and can be specialized to matrix-based libraries when necessary. This allows our development to supply the main analytic component needed to fill the corresponding gap in the existing generalized quantum Stein's lemma formalization, making it possible to complete that Lean formalization of the generalized quantum Stein's lemma without relying on the corresponding \texttt{sorry}~\cite{MeibergPrivate2026}. In this sense, the present work contributes not only an isolated formal theorem but also a bridge between abstract operator-theoretic formalization and existing quantum-information libraries.

Beyond these applications, the codebase developed here can support formalizations beyond the sandwiched R\'enyi relative entropy. Operator monotonicity, generalized perspectives, Lieb--Ando inequalities, Young inequalities, tensor-product CFC identities, and Haar-unitary averaging are common tools. Once formalized with stable interfaces, they can be reused for converse bounds, quantum hypothesis testing, resource theories, and other operator-inequality arguments.

\subsection{Organization of the Paper}

The rest of this paper is organized as follows.
Section~\ref{sec:qm} introduces the formalization of finite-dimensional quantum systems and channels. Section~\ref{sec:trace} develops the operator and trace inequalities used in the proof. 
Section~\ref{sec:entropy} formalizes the sandwiched R\'enyi relative entropy and uses the library infrastructure to prove the DPI as a key demonstration.
Section~\ref{sec:conclusion} concludes with the current scope and future directions.

\section{Basis-Independent Lean Formalization of Quantum Mechanics}
\label{sec:qm}

This section describes the finite-dimensional quantum-mechanical layer of the formalization. Its role is to provide a coordinate-free interface for quantum systems, states, and channels that can interact smoothly with the operator-theoretic infrastructure of \textit{Mathlib}. Quantum systems are represented as finite-dimensional complex Hilbert spaces, and quantum operations are represented as linear maps between operator algebras, rather than as matrices with a fixed index type. Basis-dependent constructions are introduced only where they are mathematically necessary, while the primitive notions of system, operator, state, positivity, trace preservation, and complete positivity remain basis independent.

The section is organized as follows. In Sec.\ref{subsec:states}, we define quantum systems, operator spaces, traces, and the basic predicates for positivity, positive definiteness, projections, and density operators. In Sec.\ref{subsec:channels}, we define superoperators and quantum channels, and we describe the algebraic infrastructure connecting linear endomorphisms with the continuous-linear-map and $C^\ast$-algebraic interfaces used by \textit{Mathlib}. We then describe tensor-product systems, partial traces, vectorization, basis-dependent transposition, Choi operators, Kraus representations, Stinespring representations, and Ando's identity. These constructions provide the finite-dimensional quantum-information interface used in the later trace-inequality and entropy sections.

\subsection{Quantum Systems and States}
\label{subsec:states}

A finite-dimensional quantum system is represented by a Lean type equipped with the structures of a complex Hilbert space and finite dimensionality:
\begin{lstlisting}[language=Lean]
class Qudit (a : Type u) extends
  NormedAddCommGroup a,
  InnerProductSpace ℂ a,
  CompleteSpace a,
  FiniteDimensional ℂ a

abbrev L (ℋ : Type u) [AddCommGroup ℋ] [Module ℂ ℋ] : Type u :=
  ℋ →ₗ[ℂ] ℋ

noncomputable abbrev Tr : L ℋ →ₗ[ℂ] ℂ := LinearMap.trace ℂ ℋ
\end{lstlisting}
Mathematically, an instance of \texttt{Qudit} is a finite-dimensional complex Hilbert space $\mathcal H$, and \texttt{L H} is the algebra $L(\mathcal H)$ of complex-linear endomorphisms. The trace is the coordinate-free trace from \textit{Mathlib}, rather than a sum over a distinguished standard basis.

The choice of linear endomorphisms is convenient for algebraic manipulations in finite-dimensional quantum information theory. At the same time, many analytic parts of \textit{Mathlib} are formulated for continuous linear maps. Since every linear map on a finite-dimensional normed space is continuous, the formalization can move between these representations when needed. This bridge is used repeatedly in later sections, where algebraic expressions involving quantum states and channels must interact with spectra, positivity, the continuous functional calculus, and $C^\ast$-algebraic structure.

The basic predicates for operators are defined in the ambient operator space:
\begin{lstlisting}[language=Lean]
def IsPositiveDefinite (X : L ℋ) : Prop :=
  X.IsPositive ∧ X.det ≠ 0

def IsProjective (X : L ℋ) : Prop :=
  X.IsPositive ∧ IsIdempotentElem X

def IsDensity (X : L ℋ) : Prop :=
  X.IsPositive ∧ Tr X = 1
\end{lstlisting}
Thus, a density operator is an operator $X\in L(\mathcal H)$ satisfying $X\ge0$ and $\Tr[X]=1$. A projective operator is a positive idempotent, and a positive definite operator is represented by positivity together with nonvanishing determinant, equivalently invertibility in finite dimension. In later sections, strict positivity is also expressed through spectral-domain conditions, such as $\operatorname{Spec}[X]\subset(0,\infty)$, depending on which interface is needed for the continuous functional calculus. This distinction matters because positivity is sufficient for density operators and many order-theoretic statements, whereas strict positivity or invertibility is required for inverse powers, inverse square roots, generalized perspectives, and the positive definite version of the sandwiched R\'enyi relative entropy.

These definitions implement the predicate-based design described in the introduction. Rather than introducing separate bundled types for density operators, projections, or positive definite operators, we state the relevant conditions as hypotheses on ambient operators. For example, a typical theorem fixes $\rho,\sigma\in L(\mathcal H)$ and assumes
\begin{align}
    \rho\ge0,\qquad
    \sigma>0,\qquad
    \Tr[\rho]=1,
\end{align}
rather than quantifying over several nested subtypes. This avoids repeated coercions and makes it easier to apply library results stated for ambient objects satisfying order, spectral, or algebraic hypotheses.

The use of type classes also avoids fixing a concrete model such as $\mathbb C^d$. A theorem can quantify directly over an arbitrary finite-dimensional Hilbert space $\mathcal H$ with a \texttt{Qudit} instance. Basis choices therefore occur only when they are mathematically part of the construction, for example, in vectorization, transposition, Choi representations, or spectral decompositions. The definitions of systems, states, channels, positivity, and trace preservation remain invariant under changes of basis.

This differs from matrix-first presentations such as Carlen's introductory course, where the starting point is the matrix algebra $M_n$ and its Hermitian part $\mathcal H_n$\cite{Carlen2010}. The matrix viewpoint is natural for exposition, but it is less convenient for a library intended to interact with \textit{Mathlib}'s abstract operator theory. Our formulation internalizes finite dimensionality as type-class data and treats matrices as coordinate representations introduced only when needed. This does not change the underlying finite-dimensional mathematics; it changes the interface so that statements are coordinate-free, dimension bookkeeping is reduced, and existing algebraic and analytic APIs can be reused more directly.

\subsection{Quantum Channels and Channel Representations}
\label{subsec:channels}

Quantum operations are first represented as linear maps between operator spaces:
\begin{lstlisting}[language=Lean]
abbrev T (ℋ₁ : Type u) (ℋ₂ : Type v)
  [AddCommGroup ℋ₁] [Module ℂ ℋ₁] [AddCommGroup ℋ₂] [Module ℂ ℋ₂] :
  Type (max u v) :=
  (L ℋ₁) →ₗ[ℂ] (L ℋ₂)
\end{lstlisting}
Thus \texttt{T H1 H2} is the type of superoperators from $L(\mathcal H_1)$ to $L(\mathcal H_2)$. Complete positivity is expressed by comparison with \textit{Mathlib}'s bundled completely positive maps:
\begin{lstlisting}[language=Lean]
def IsCompletelyPositive (Φ : T ℋ₁ ℋ₂) : Prop :=
  ∃ Ψ : CompletelyPositiveMap (L ℋ₁) (L ℋ₂),
    Ψ.toLinearMap = Φ
\end{lstlisting}
This definition is deliberately tied to the existing \textit{Mathlib} notion of complete positivity, while keeping the visible quantum-information object as an unbundled linear superoperator. For trace-preserving quantum channels, the code also defines a bundled structure:
\begin{lstlisting}[language=Lean]
structure CPTP (ℋ₁ : Type u) (ℋ₂ : Type v) [Qudit ℋ₁] [Qudit ℋ₂]
  extends CompletelyPositiveMap (L ℋ₁) (L ℋ₂) where
  trace_map (ρ : L ℋ₁) : Tr ρ = Tr (toFun ρ)
\end{lstlisting}
Thus a channel
\begin{align}
    \Phi:L(\mathcal H_1)\to L(\mathcal H_2)
\end{align}
is represented as a completely positive map together with the trace-preservation condition
\begin{align}
    \Tr[\Phi(\rho)]=\Tr[\rho].
\end{align}
Kraus and Stinespring representations are not taken as primitive definitions of a channel. They are derived and related to complete positivity through separate representation theorems. This keeps the primitive notion close to the operator-algebraic definition of a quantum channel and makes it possible to use whichever representation is appropriate in a later proof.

The main technical issue is that the natural object in finite-dimensional quantum information theory is the algebra $L(\mathcal H)$ of linear endomorphisms, whereas many analytic results in \textit{Mathlib} are stated for continuous linear maps or for abstract $C^\ast$-algebras with ordered-star structure. Although these viewpoints are equivalent in finite dimension, Lean treats them as different types. The formalization therefore builds compatibility between the linear-map representation used for quantum-information expressions and the continuous-linear-map representation used by the analytic library. In particular, $L(\mathcal H)$ is equipped with normed, metric, $C^\ast$-algebraic, and ordered-star structures by transporting the corresponding structures from continuous linear maps:
\begin{lstlisting}[language=Lean]
noncomputable instance (ℋ : Type u) [Qudit ℋ] : Norm (L ℋ) where
  norm := fun X => norm X.toContinuousLinearMap

noncomputable instance (ℋ : Type u) [Qudit ℋ] : MetricSpace (L ℋ) :=
  MetricSpace.induced LinearMap.toContinuousLinearMap
    LinearMap.toContinuousLinearMap.injective inferInstance

noncomputable instance (ℋ : Type u) [Qudit ℋ] : CStarAlgebra (L ℋ) where
  dist_eq x y := continuous_instance.dist_eq
    x.toContinuousLinearMap y.toContinuousLinearMap
  norm_mul_le x y := continuous_instance.norm_mul_le
    x.toContinuousLinearMap y.toContinuousLinearMap
  complete := (linear_isometry_equiv.completeSpace_iff.mpr
    continuous_instance.toCompleteSpace).complete
  norm_mul_self_le x := continuous_instance.norm_mul_self_le x.toContinuousLinearMap
  algebraMap := c_algebra_instance.algebraMap
  commutes' := c_algebra_instance.commutes'
  smul_def' := c_algebra_instance.smul_def'
  norm_smul_le r x := continuous_instance.norm_smul_le r x.toContinuousLinearMap

noncomputable instance (ℋ : Type u) [Qudit ℋ] : StarOrderedRing (L ℋ) where
  le_iff x y := by
    constructor
    · intro h
      obtain <p, hp, hy> :=
        (continuous_sor_instance.le_iff
          x.toContinuousLinearMap y.toContinuousLinearMap).mp h
      use p.toLinearMap
      exact <closure_lm_of_closure_clm hp,
        LinearMap.toContinuousLinearMap.injective (hy.trans rfl)>
    · rintro <p, hp, hy>
      apply (continuous_sor_instance.le_iff
        x.toContinuousLinearMap y.toContinuousLinearMap).mpr
      exact <p.toContinuousLinearMap, closure_clm_of_closure_lm hp,
        congrArg LinearMap.toContinuousLinearMap hy>
\end{lstlisting}
This transportation step allows later arguments to use \textit{Mathlib}'s infrastructure for spectra, positivity, continuous functional calculus, and completely positive maps while retaining the simpler algebraic notation of linear endomorphisms. Without this bridge, many proofs would require repeated translations among matrix-style calculations, linear maps, and continuous linear maps.

Tensor-product systems are also made instances of the same quantum-system class. This is necessary because channels, partial traces, Choi operators, and Stinespring representations all involve composite systems. The formalization treats $\mathcal H_1\otimes\mathcal H_2$ as another finite-dimensional Hilbert space:
\begin{lstlisting}[language=Lean]
noncomputable instance : Qudit (ℋ₁ ⊗[ℂ] ℋ₂) := by
  letI : Module.Finite ℂ (ℋ₁ ⊗[ℂ] ℋ₂) := inferInstance
  exact
    { toNormedAddCommGroup := inferInstance
      toInnerProductSpace := inferInstance
      toCompleteSpace := inferInstance
      fg_top := Module.Finite.fg_top }
\end{lstlisting}
Consequently, $L(\mathcal H_1\otimes\mathcal H_2)$ can be used with the same algebraic and analytic interfaces as $L(\mathcal H)$.

A recurring construction is the equivalence between endomorphisms of a tensor-product Hilbert space and tensor products of endomorphism spaces. On simple tensors, the main compatibility lemmas are:
\begin{lstlisting}[language=Lean]
lemma l_tensor_equiv_symm_tmul
    {E : Type u} {F : Type v} [Qudit E] [Qudit F]
    (B : L E) (C : L F) :
    (l_tensor_equiv (ℋ₁ := E) (ℋ₂ := F)).symm (B ⊗ₜ[ℂ] C) =
      TensorProduct.map B C

lemma l_tensor_equiv_map_tmul
    {E : Type u} {F : Type v} [Qudit E] [Qudit F]
    (B : L E) (C : L F) :
    (l_tensor_equiv (ℋ₁ := E) (ℋ₂ := F)) (TensorProduct.map B C) =
      B ⊗ₜ[ℂ] C
\end{lstlisting}
These lemmas are used repeatedly in the definitions and computations of partial traces, Choi operators, Kraus expansions, and Stinespring representations.

The partial trace is implemented by combining this tensor-hom equivalence with the ordinary trace. One of the trace-out maps used in the code is:
\begin{lstlisting}[language=Lean]
noncomputable def Tr₂ : T (ℋ₁ ⊗[ℂ] ℋ₂) ℋ₂ :=
  (TensorProduct.lid ℂ (L ℋ₂)).toLinearMap
  ∘ₗ (TensorProduct.map Tr LinearMap.id)
  ∘ₗ l_tensor_equiv.toLinearMap
\end{lstlisting}
With the displayed type, \texttt{Tr2} maps $L(\mathcal H_1\otimes\mathcal H_2)$ to $L(\mathcal H_2)$, and hence traces out the first tensor factor. It is characterized on simple tensors by
\begin{lstlisting}[language=Lean]
lemma Tr₂_l_tensor_equiv_symm_tmul
    (X : L ℋ₁) (Y : L ℋ₂) :
    Tr₂ ((l_tensor_equiv (ℋ₁ := ℋ₁) (ℋ₂ := ℋ₂)).symm (X ⊗ₜ[ℂ] Y)) =
      (Tr X) • Y
\end{lstlisting}
Mathematically, this says that
\begin{align}
    \operatorname{Tr}_1(X\otimes Y)=\Tr[X]Y.
\end{align}
For Stinespring representations, the code also uses a right partial trace, defined by conjugating with the tensor-factor swap and then applying \texttt{Tr2}:
\begin{lstlisting}[language=Lean]
noncomputable def TrRight {ℋ₃ : Type u} [Qudit ℋ₃] :
  T (ℋ₂ ⊗[ℂ] ℋ₃) ℋ₂ :=
  (Tr₂ (ℋ₁ := ℋ₃) (ℋ₂ := ℋ₂)).comp
    (conjugateEnd (TensorProduct.comm ℂ ℋ₂ ℋ₃))
\end{lstlisting}
This is the map that traces out the second tensor factor in $L(\mathcal H_2\otimes\mathcal H_3)$.

Vectorization is the first point in this section where a basis choice is mathematically necessary. The code therefore makes this dependence explicit:
\begin{lstlisting}[language=Lean]
noncomputable def vec (b : Module.Basis i ℂ ℋ₂) :
  (ℋ₂ →ₗ[ℂ] ℋ₁) →ₗ[ℂ] (ℋ₁ ⊗[ℂ] ℋ₂) :=
  (TensorProduct.comm ℂ ℋ₂ ℋ₁)
  ∘ₗ (TensorProduct.map b.toDualEquiv.symm.toLinearMap LinearMap.id)
  ∘ₗ (dualTensorHomEquiv ℂ ℋ₂ ℋ₁).symm.toLinearMap
\end{lstlisting}
If $b=(e_i)_i$ is an orthonormal basis of $\mathcal H_2$, this map sends a linear operator $K:\mathcal H_2\to\mathcal H_1$ to
\begin{align}
    \operatorname{vec}_b(K)=\sum_i K e_i\otimes e_i.
\end{align}
The basic compatibility with the Hilbert--Schmidt inner product is:
\begin{lstlisting}[language=Lean]
lemma inner_vec_eq_trace
  (b : OrthonormalBasis i ℂ ℋ₂)
  (A B : ℋ₂ →ₗ[ℂ] ℋ₁) :
  inner ℂ (vec b.toBasis A) (vec b.toBasis B) =
    Tr (A† ∘ₗ B)
\end{lstlisting}
That is,
\begin{align}
    \langle \operatorname{vec}_b(A),\operatorname{vec}_b(B)\rangle
    =
    \Tr[A^\dagger B].
\end{align}
This lemma is used to move between tensor-product inner products and trace expressions.

The same basis dependence requires a basis-dependent transpose. The formalization defines the transpose of an operator relative to the same basis used for vectorization:
\begin{lstlisting}[language=Lean]
noncomputable def l_transpose (b : Module.Basis i ℂ ℋ₁)
  (A : L ℋ₁) : L ℋ₁ :=
  b.toDualEquiv.symm.toLinearMap ∘ₗ A.dualMap ∘ₗ b.toDualEquiv.toLinearMap
\end{lstlisting}
This transpose is not treated as a canonical basis-independent operation, because transposition is not intrinsic on an abstract Hilbert space. Making the basis dependence explicit prevents an ambiguity common in informal tensor-calculus notation.

Using vectorization and this transpose, the formalization proves Ando's identity:
\begin{lstlisting}[language=Lean]
lemma ando_identity (b : OrthonormalBasis i ℂ ℋ₂)
  (A : L ℋ₁) (B : L ℋ₂) (K : ℋ₂ →ₗ[ℂ] ℋ₁) :
  inner ℂ (vec b.toBasis K) ((TensorProduct.map A B) (vec b.toBasis K))
    = Tr (K† ∘ₗ A ∘ₗ K ∘ₗ (l_transpose b.toBasis B))
\end{lstlisting}
Equivalently,
\begin{align}
    \left\langle
        \operatorname{vec}_b(K),
        (A\otimes B)\operatorname{vec}_b(K)
    \right\rangle
    =
    \Tr\left[K^\dagger A K B^{\mathsf T_b}\right],
\end{align}
where $B^{\mathsf T_b}$ denotes the transpose of $B$ with respect to $b$. This identity connects tensor-product positivity with trace expressions and is used later in the trace-inequality infrastructure. Carlen's course uses the corresponding matrix identity in the development of trace inequalities~\cite{Carlen2010}; here the same identity is proved for abstract finite-dimensional Hilbert spaces, with the basis dependence isolated in the vectorization and transpose constructions.

The vectorization machinery is also used to define Choi operators. The code first defines rank-one operators by the dual-tensor equivalence:
\begin{lstlisting}[language=Lean]
noncomputable def outer_product
  (u : ℋ₁) (v : ℋ₂) : ℋ₁ →ₗ[ℂ] ℋ₂ :=
  (dualTensorHomEquiv ℂ ℋ₁ ℋ₂).toLinearMap <|
    ((InnerProductSpace.toDualMap ℂ ℋ₁ u) ⊗ₜ[ℂ] v)
\end{lstlisting}
Given a superoperator $\Phi:T(\mathcal H_1,\mathcal H_2)$ and a basis of $\mathcal H_1$, the Choi operator is defined by applying the tensor extension of $\Phi$ to the rank-one operator associated with the vectorized identity:
\begin{lstlisting}[language=Lean]
noncomputable def choi (b : Module.Basis i ℂ ℋ₁) (Φ : T ℋ₁ ℋ₂) :
  L (ℋ₂ ⊗[ℂ] ℋ₁) :=
  (l_tensor_equiv.symm.toLinearMap
  ∘ₗ (TensorProduct.map Φ LinearMap.id)
  ∘ₗ l_tensor_equiv.toLinearMap)
    (outer_product (vec b (I ℋ₁)) (vec b (I ℋ₁)))

def ChoiPositive (b : Module.Basis i ℂ ℋ₁) (Φ : T ℋ₁ ℋ₂) : Prop :=
  0 ≤ choi b Φ
\end{lstlisting}
Mathematically, this is the Choi operator
\begin{align}
    C_\Phi =
    (\Phi\otimes\id)
    \left[
        \operatorname{vec}_b(I)\operatorname{vec}_b(I)^\dagger
    \right],
\end{align}
up to the tensor-factor convention fixed by the code.

The tensor extension used in this comparison is encoded by:
\begin{lstlisting}[language=Lean]
noncomputable def amplifyWithId (Φ : T ℋ₁ ℋ₂) :
  T (ℋ₁ ⊗[ℂ] ℋ₁) (ℋ₂ ⊗[ℂ] ℋ₁) :=
  l_tensor_equiv.symm.toLinearMap
    ∘ₗ TensorProduct.map Φ LinearMap.id
    ∘ₗ l_tensor_equiv.toLinearMap

def IsPositiveMap (Φ : T ℋ₁ ℋ₂) : Prop :=
  ∀ X : L ℋ₁, 0 ≤ X → 0 ≤ Φ X

def AmplificationPositive (Φ : T ℋ₁ ℋ₂) : Prop :=
  IsPositiveMap (amplifyWithId Φ)
\end{lstlisting}
Thus \texttt{AmplificationPositive Phi} states that the tensor extension of $\Phi$ with the identity map is positive.

A central point of this channel layer is the formalization of the standard finite-dimensional representation theorem for complete positivity. Instead of encoding the result only as a single large theorem, the code defines the relevant conditions as separate predicates and proves the implications among them. Kraus representations are defined with an arbitrary finite index type:
\begin{lstlisting}[language=Lean]
def KrausRep (Φ : T ℋ₁ ℋ₂) (κ : Type*) [DecidableEq κ] [Fintype κ] :
  Prop :=
  ∃ A : κ → (ℋ₁ →ₗ[ℂ] ℋ₂),
    ∀ X : L ℋ₁, Φ X = ∑ a : κ, (A a).comp (X.comp (LinearMap.adjoint (A a)))

def HasKraus (Φ : T ℋ₁ ℋ₂) : Prop :=
  ∃ (κ : Type u) (_ : DecidableEq κ) (_ : Fintype κ),
    @KrausRep ℋ₁ ℋ₂ _ _ Φ κ _ _
\end{lstlisting}
Mathematically, \texttt{KrausRep Phi k} asserts that
\begin{align}
    \Phi(X)=\sum_{a\in\kappa} A_a X A_a^\dagger.
\end{align}
The code also defines a rank-bounded Kraus representation using the rank of the Choi operator:
\begin{lstlisting}[language=Lean]
noncomputable def choiRank (b : Module.Basis i ℂ ℋ₁) (Φ : T ℋ₁ ℋ₂) : Nat :=
  Module.finrank ℂ (LinearMap.range (choi b Φ))

def HasRankKraus (b : Module.Basis i ℂ ℋ₁) (Φ : T ℋ₁ ℋ₂) : Prop :=
  ∃ (κ : Type u) (dec : DecidableEq κ) (inst : Fintype κ),
    Fintype.card κ = choiRank b Φ ∧
      @KrausRep ℋ₁ ℋ₂ _ _ Φ κ dec inst
\end{lstlisting}

Stinespring representations are defined using an auxiliary finite-dimensional Hilbert space:
\begin{lstlisting}[language=Lean]
def StinespringRep (Φ : T ℋ₁ ℋ₂) (ℋ₃ : Type u) [Qudit ℋ₃] : Prop :=
  ∃ A : ℋ₁ →ₗ[ℂ] (ℋ₂ ⊗[ℂ] ℋ₃),
    ∀ X : L ℋ₁,
      Φ X =
        (@TrRight ℋ₂ inferInstance ℋ₃ inferInstance
          (((A.comp X).comp
            (LinearMap.adjoint A : (ℋ₂ ⊗[ℂ] ℋ₃) →ₗ[ℂ] ℋ₁)) :
              L (ℋ₂ ⊗[ℂ] ℋ₃)) : L ℋ₂)

def HasStinespring (Φ : T ℋ₁ ℋ₂) : Prop :=
  ∃ (ℋ₃ : Type u) (_ : Qudit ℋ₃),
    @StinespringRep ℋ₁ ℋ₂ _ _ Φ ℋ₃ _
\end{lstlisting}
Thus the formal statement corresponds to
\begin{align}
    \Phi(X)=\operatorname{Tr}_E[A X A^\dagger]
\end{align}
for some linear map $A:\mathcal H_1\to\mathcal H_2\otimes E$. The rank-bounded version requires the environment dimension to equal the Choi rank:
\begin{lstlisting}[language=Lean]
def HasRankStinespring (b : Module.Basis i ℂ ℋ₁) (Φ : T ℋ₁ ℋ₂) : Prop :=
  ∃ (ℋ₃ : Type u) (inst : Qudit ℋ₃),
    Module.finrank ℂ ℋ₃ = choiRank b Φ ∧
      @StinespringRep ℋ₁ ℋ₂ _ _ Φ ℋ₃ inst
\end{lstlisting}

The main equivalence theorem relates complete positivity, positivity of tensor ampliations, Choi positivity, Kraus representations, rank Kraus representations, Stinespring representations, and rank Stinespring representations. In the code, the proof is decomposed into named implications:
\begin{lstlisting}[language=Lean]
theorem cp_to_tensor
    (Φ : T ℋ₁ ℋ₂) :
    IsCompletelyPositive Φ → AmplificationPositive Φ

theorem tensor_to_choi
    (b : Module.Basis i ℂ ℋ₁) (Φ : T ℋ₁ ℋ₂) :
    AmplificationPositive Φ → ChoiPositive b Φ

theorem choi_to_rank_kraus
    (b : Module.Basis i ℂ ℋ₁) (Φ : T ℋ₁ ℋ₂) :
    ChoiPositive b Φ → HasRankKraus b Φ

theorem kraus_to_stinespring
    (Φ : T ℋ₁ ℋ₂) :
    HasKraus Φ → HasStinespring Φ

theorem rank_kraus_to_rank_stinespring
    (b : Module.Basis i ℂ ℋ₁) (Φ : T ℋ₁ ℋ₂) :
    HasRankKraus b Φ → HasRankStinespring b Φ

theorem stinespring_to_cp
    (Φ : T ℋ₁ ℋ₂) :
    HasStinespring Φ → IsCompletelyPositive Φ
\end{lstlisting}
These implications are then assembled into equivalence theorems:
\begin{lstlisting}[language=Lean]
theorem cp_iff_tensor
    (b : Module.Basis k ℂ ℋ₁) (Φ : T ℋ₁ ℋ₂) :
    IsCompletelyPositive Φ ↔ AmplificationPositive Φ

theorem cp_iff_choi
    (b : Module.Basis k ℂ ℋ₁) (Φ : T ℋ₁ ℋ₂) :
    IsCompletelyPositive Φ ↔ ChoiPositive b Φ

theorem cp_iff_kraus
    (b : Module.Basis k ℂ ℋ₁) (Φ : T ℋ₁ ℋ₂) :
    IsCompletelyPositive Φ ↔ HasKraus Φ

theorem cp_iff_rank_kraus
    (b : Module.Basis k ℂ ℋ₁) (Φ : T ℋ₁ ℋ₂) :
    IsCompletelyPositive Φ ↔ HasRankKraus b Φ

theorem cp_iff_stinespring
    (b : Module.Basis k ℂ ℋ₁) (Φ : T ℋ₁ ℋ₂) :
    IsCompletelyPositive Φ ↔ HasStinespring Φ

theorem cp_iff_rank_stinespring
    (b : Module.Basis k ℂ ℋ₁) (Φ : T ℋ₁ ℋ₂) :
    IsCompletelyPositive Φ ↔ HasRankStinespring b Φ
\end{lstlisting}
This decomposition is a formalization choice rather than a mathematical change. Different later arguments require different directions of the equivalence. The proof of the DPI uses Stinespring dilation and trace preservation, while other applications may use Choi positivity or Kraus operators. Providing each implication separately avoids repeatedly unpacking a large equivalence theorem and gives stable intermediate targets for proof development.

The proof of the equivalence also differs in organization from a purely matrix-first treatment. Since \textit{Mathlib}'s complete positivity interface is closely tied to positivity of matrix amplifications, the code constructs an explicit bridge between matrix-style ampliations and tensor ampliations. It introduces finite direct sums, matrix algebras over $L(\mathcal H)$, and the corresponding equivalence with operators on tensor-product Hilbert spaces. This allows the proof to move from complete positivity in the \textit{Mathlib} sense to positivity of the tensor extension $\Phi\otimes\id$, and then to Choi positivity by applying the tensor extension to $\operatorname{vec}_b(I)\operatorname{vec}_b(I)^\dagger$. Conversely, Choi positivity is converted to a rank Kraus representation by decomposing a positive Choi operator into a finite sum of outer products and then recovering Kraus operators through vectorization. Kraus operators are bundled into a Stinespring operator, and Stinespring representations imply complete positivity because the right partial trace is itself shown to be completely positive through a Kraus expansion using slice maps.

The resulting channel layer separates coordinate-free primitives from basis-dependent representation tools. States, channels, positivity, trace preservation, and complete positivity are formulated at the level of abstract finite-dimensional Hilbert spaces and their operator algebras. Basis-dependent constructions such as vectorization, transpose, and Choi operators are introduced only when needed. This separation is used throughout the rest of the paper: the analytic inequalities are formulated in operator-theoretic language, while the finite-dimensional representation theorems supply the channel and dilation tools required for the DPI.

\section{Hierarchy of Trace Inequalities}
\label{sec:trace}

This section presents the hierarchy of operator and trace inequalities that forms the central part of the library and supplies the tools used in the proof of the DPI.
The hierarchy is organized through a route of using perspective functions. Starting from operator monotonicity and convexity results for real powers, we formalize Jensen's operator inequality, generalized perspectives, operator power means, and finally the Lieb--Ando trace inequalities needed for the sandwiched R\'enyi relative entropy.

The main design choice is not to formalize only the Lieb--Ando trace inequalities in isolation. Instead, the development builds reusable interfaces for the operator-theoretic mechanisms behind them. Frank and Lieb use Lieb's concavity theorem and Ando's convexity theorem as established analytic inputs in their proof of the DPI~\cite{FrankLieb2013}. Carlen's notes present much of the surrounding operator-inequality technology in finite-dimensional matrix language~\cite{Carlen2010}. Our formalization follows the same mathematical tradition but reorganizes the proof around generalized perspectives and operator power means, following the simpler proof strategy of Ref.~\cite{NIKOUFAR2013531}. In this route, Jensen's operator inequality implies joint convexity and concavity of generalized perspectives; generalized perspectives yield operator power means; and operator power means, applied to left and right multiplication on Hilbert--Schmidt operator spaces, yield the Lieb--Ando trace inequalities.

This organization has two formal advantages. First, it concentrates the main proof obligations, including spectral-domain bookkeeping, inverse square roots, contraction weights, Jensen-type dilations, positivity of block expressions, and compatibility with the continuous functional calculus, into reusable intermediate theorems. Second, it avoids repeatedly translating informal matrix identifications into Lean. Classical proofs often identify a block matrix with an operator on a direct sum, or identify $L(\mathcal H)$ with a Hilbert space under the Hilbert--Schmidt inner product. In Lean, these identifications must be represented by definitions, equivalences, coercions, and compatibility lemmas. Once these interfaces are built, the trace inequalities become applications of already formalized statements on operators rather than separate matrix-level calculations.

The section proceeds as follows. In Sec.~\ref{subsec:lowner}, we define operator monotonicity, antitonicity, convexity, and concavity through the real continuous functional calculus, and package the real-power results used later. In Sec.~\ref{subsec:block}, we describe the block-operator and matrix-positivity infrastructure used in Schur-complement and dilation arguments. In Sec.~\ref{subsec:hs}, we turn the operator space into a Hilbert space and formalize left and right multiplication, including their compatibility with the continuous functional calculus. In Sec.~\ref{subsec:jensen}, we formalize Jensen's operator inequality in the contraction and two-operator forms needed for perspectives. In Sec.~\ref{subsec:perspective}, we prove joint convexity and concavity of generalized perspectives. In Sec.~\ref{subsec:mean}, we specialize this theorem to operator power means. Finally, in Sec.~\ref{subsec:lieb}, we derive the Lieb--Ando trace inequalities.

\subsection{Operator Monotonicity and L\"owner--Heinz Theorem}
\label{subsec:lowner}

The first layer is a uniform interface for operator monotonicity, antitonicity, convexity, and concavity.
We first formalized the main proofs, where the ambient object is an abstract ordered $C^\ast$-algebra equipped with real continuous functional calculus.
Then, we further specialize these results to bounded operators $L(\mathcal H)$ on a complex Hilbert space. This separation is important: the core results are not tied to finite-dimensional Hilbert spaces, while the quantum-information arguments instantiate this bounded-operator wrapper in finite-dimensional settings.

The basic real continuous functional calculus notation is introduced as:
\begin{lstlisting}[language=Lean]
noncomputable abbrev cfcR (f : ℝ → ℝ) (A : Alg) : Alg :=
  cfc (R := ℝ) (A := Alg) (p := IsSelfAdjoint) f A
\end{lstlisting}
Here \texttt{Alg} denotes the ambient $C^\ast$-algebra in the Core file. This is not a new functional calculus; it is a local abbreviation for \textit{Mathlib}'s real continuous functional calculus applied to self-adjoint elements.

The core predicates are formulated with explicit spectral-domain conditions. A representative part of the interface is:
\begin{lstlisting}[language=Lean]
def OperatorMonotoneOn (s : Set ℝ) (f : ℝ → ℝ) : Prop :=
  ∀ {A B : Alg},
    0 ≤ A → 0 ≤ B → B ≤ A →
    spectrum ℝ A ⊆ s → spectrum ℝ B ⊆ s →
    cfcR f B ≤ cfcR f A

def OperatorAntitoneOn (s : Set ℝ) (f : ℝ → ℝ) : Prop :=
  ∀ {A B : Alg},
    0 ≤ A → 0 ≤ B → B ≤ A →
    spectrum ℝ A ⊆ s → spectrum ℝ B ⊆ s →
    cfcR f A ≤ cfcR f B

def OperatorConvexOn (s : Set ℝ) (f : ℝ → ℝ) : Prop :=
  ∀ {A B : Alg} {t : ℝ},
    IsSelfAdjoint A → IsSelfAdjoint B →
    0 ≤ t → t ≤ 1 →
    spectrum ℝ A ⊆ s → spectrum ℝ B ⊆ s →
    cfcR f ((1 - t) • A + t • B)
      ≤ (1 - t) • cfcR f A + t • cfcR f B

def OperatorConcaveOn (s : Set ℝ) (f : ℝ → ℝ) : Prop :=
  OperatorConvexOn s (fun x => - f x)
\end{lstlisting}
The exact type-class assumptions are supplied in the implementation; in particular, the Core theorem section uses \texttt{NonnegSpectrumClass R Alg} to connect order positivity with spectral nonnegativity. The important point is that monotonicity and antitonicity compare positive self-adjoint elements with spectral restrictions, while convexity and concavity are stated for self-adjoint elements whose spectra lie in the prescribed set. This distinction is essential in Lean because the real continuous functional calculus is only applied after the relevant self-adjointness and spectral-domain hypotheses have been discharged.

The code also introduces uniform versions, such as \texttt{OperatorConvexOnAll}. In the Core file, these quantify over all $C^\ast$-algebras with the required structures. In the Hilbert-space wrapper, the corresponding uniform predicates quantify over Hilbert spaces. This distinction matters because Jensen's inequality and the power-mean arguments are applied not only to the original Hilbert space, but also to block spaces and Hilbert--Schmidt operator spaces. The uniform predicates avoid restating scalar function properties after every change of ambient space.

The proof hierarchy begins with the inverse function. The antitonicity of $x\mapsto 1/x$ on $(0,\infty)$ is packaged as:
\begin{lstlisting}[language=Lean]
theorem one_div_operatorAntitoneOn_Ioi :
  OperatorAntitoneOn (Set.Ioi (0 : ℝ)) (fun x : ℝ => 1 / x)
\end{lstlisting}
Mathematically, this says that if $0<B\le A$ and both spectra lie in $(0,\infty)$, then
\begin{align}
    A^{-1}\le B^{-1}.
\end{align}
The Lean proof treats the inverse as a continuous-functional-calculus expression and uses the $C^\ast$-algebraic inverse-order API for positive invertible elements. Thus the proof is not written as a finite-dimensional matrix argument. The spectral condition $\operatorname{Spec}[A]\subset(0,\infty)$ is an explicit hypothesis because positivity alone does not justify applying the inverse function.

The convexity of the inverse is also formalized:
\begin{lstlisting}[language=Lean]
theorem one_div_operatorConvexOn_Ioi :
  OperatorConvexOn (Set.Ioi (0 : ℝ)) (fun x : ℝ => 1 / x)
\end{lstlisting}
The proof follows the Schur-complement idea, but in the Core file it is formulated using $2\times2$ matrices with entries in the ambient $C^\ast$-algebra and the predicate \texttt{Matrix.PosSemidef}. This is slightly different from the usual presentation in Carlen's notes, where the argument is written for block matrices over finite-dimensional matrix algebras. In the formal proof, the block matrix
\begin{align}
    \begin{pmatrix}
        A & 1\\
        1 & A^{-1}
    \end{pmatrix}
\end{align}
is treated as an element of a matrix algebra over the ambient $C^\ast$-algebra. Positivity is preserved under convex combinations and congruence by invertible matrices, and the Schur-complement conclusion gives the operator convexity of the inverse.

The next step is the shifted inverse. For $t>0$, the code proves operator antitonicity and convexity of
\begin{align}
    x\mapsto \frac{1}{x+t}
\end{align}
on $[0,\infty)$:
\begin{lstlisting}[language=Lean]
theorem one_div_add_t_operatorAntitoneOn_Ici :
  ∀ t : ℝ, 0 < t →
    OperatorAntitoneOn (Set.Ici (0 : ℝ)) (fun x : ℝ => 1 / (x + t))

theorem one_div_add_t_operatorConvexOn_Ici :
  ∀ t : ℝ, 0 < t →
    OperatorConvexOn (Set.Ici (0 : ℝ)) (fun x : ℝ => 1 / (x + t))
\end{lstlisting}
This intermediate layer is important in the current codebase. The ratio function is then derived from the shifted inverse using
\begin{align}
    \frac{x}{x+t}
    =
    1-\frac{t}{x+t}.
\end{align}
The corresponding interface is:
\begin{lstlisting}[language=Lean]
theorem ratio_add_t_operatorMonotoneOn_Ici :
  ∀ t : ℝ, 0 < t →
    OperatorMonotoneOn (Set.Ici (0 : ℝ)) (fun x : ℝ => x / (x + t))

theorem ratio_add_t_operatorConcaveOn_Ici :
  ∀ t : ℝ, 0 < t →
    OperatorConcaveOn (Set.Ici (0 : ℝ)) (fun x : ℝ => x / (x + t))
\end{lstlisting}
This organization reflects the implementation rather than only the natural-language proof. The shifted inverse isolates the spectral shift from $[0,\infty)$ to $(0,\infty)$, and the ratio function then follows by affine operations and sign reversal.

The power-function results are packaged as the L\"owner--Heinz interface. The positive-power results include:
\begin{lstlisting}[language=Lean]
theorem power_Icc_zero_one_operatorMonotoneOn_Ici : ∀ p ∈ Set.Icc (0 : ℝ) 1,
    OperatorMonotoneOn (Set.Ici (0 : ℝ)) (fun x : ℝ => x ^ p)

theorem power_Icc_zero_one_operatorConcaveOn_Ici : ∀ p ∈ Set.Icc (0 : ℝ) 1,
    OperatorConcaveOn (Set.Ici (0 : ℝ)) (fun x : ℝ => x ^ p)

theorem power_Icc_one_two_operatorConvexOn_Ici : ∀ p ∈ Set.Icc (1 : ℝ) 2,
    OperatorConvexOn (Set.Ici (0 : ℝ)) (fun x : ℝ => x ^ p)
\end{lstlisting}
These express the standard facts that $x^p$ is operator monotone and operator concave on $[0,\infty)$ for $0\le p\le1$, and operator convex on $[0,\infty)$ for $1\le p\le2$. The monotonicity statement uses \textit{Mathlib}'s CFC power monotonicity interface. The concavity and convexity proofs are not merely restatements of one existing theorem; they combine the shifted-inverse and ratio inequalities with \textit{Mathlib}'s abstract integral representation for real powers in the continuous functional calculus.

The current Core interface also includes negative-power results, which are needed later in the Lieb--Ando trace inequalities:
\begin{lstlisting}[language=Lean]
theorem power_Icc_neg_one_zero_neg_operatorMonotoneOn_Ioi : ∀ p ∈ Set.Icc (-1 : ℝ) 0,
    OperatorMonotoneOn (Set.Ioi (0 : ℝ)) (fun x : ℝ => - (x ^ p))

theorem power_Icc_neg_one_zero_neg_operatorConcaveOn_Ioi : ∀ p ∈ Set.Icc (-1 : ℝ) 0,
    OperatorConcaveOn (Set.Ioi (0 : ℝ)) (fun x : ℝ => - (x ^ p))
\end{lstlisting}
Equivalently, for $-1\le p\le0$, the negative power $x^p$ is handled by rewriting $p=-r$ with $0\le r\le1$ and combining positive-power results with inverse-order and inverse-convexity facts. The formal proof treats identities such as $x^{-r}=1/x^r$ through continuous-functional-calculus composition. This is one of the places where the positive definite domain is essential: negative powers are evaluated on $(0,\infty)$, not merely on $[0,\infty)$.

Compared with the matrix-oriented literature, the mathematical statements are standard, but the formalization changes their interface. Carlen's notes present the L\"owner--Heinz theorem and the surrounding inverse and power inequalities in finite-dimensional operator language~\cite{Carlen2010}. The Core Lean development states and proves the relevant components for self-adjoint elements of abstract ordered $C^\ast$-algebras equipped with real continuous functional calculus, and then specializes them to Hilbert-space operators through a wrapper. This makes the real-power facts reusable in later sections without rebuilding spectral-domain and self-adjointness arguments.

\subsection{Block-Operator and Matrix-Positivity Formalism}
\label{subsec:block}

Several later arguments use block matrices, Schur complements, and dilation constructions. In ordinary mathematical writing, one often treats a block matrix as an operator on $\mathcal H\oplus\mathcal H$ without naming the underlying identifications. In Lean, this must be made explicit. The formalization uses two related representations, depending on the ambient theorem.

In the L\"owner--Heinz Core file, the Schur-complement part is stated using $2\times2$ matrices over a general $C^\ast$-algebra and the predicate \texttt{Matrix.PosSemidef}. This choice is necessary because the Core theorems are not restricted to bounded operators on Hilbert spaces. Thus, the block matrix appearing in the proof of inverse convexity is not first an operator on a Hilbert-space direct sum; it is a matrix with entries in the ambient $C^\ast$-algebra. This is the right abstraction for a theorem intended to live in \texttt{LownerHeinzCore.lean}.

For the Hilbert-space arguments, especially Jensen's inequality, it is also useful to represent a two-fold Hilbert direct sum explicitly. The formalization uses a finite $L^2$ product:
\begin{lstlisting}[language=Lean]
abbrev HSum (ℋ : Type u) [NormedAddCommGroup ℋ] [InnerProductSpace ℂ ℋ] :
    Type u :=
  PiLp 2 (fun _ : Fin 2 => ℋ)
\end{lstlisting}
Mathematically, this is $\mathcal H\oplus\mathcal H$ with the Hilbert-space inner product
\begin{align}
    \langle (v_0,v_1),(w_0,w_1)\rangle
    =
    \langle v_0,w_0\rangle+\langle v_1,w_1\rangle.
\end{align}
The use of \texttt{PiLp} lets the development reuse the finite-product Hilbert-space infrastructure already present in \textit{Mathlib}.

The corresponding inclusions and projections are defined as continuous linear maps, and their adjoint relations are proved once. If $I_i:\mathcal H\to\mathcal H\oplus\mathcal H$ is the $i$th inclusion and $P_i:\mathcal H\oplus\mathcal H\to\mathcal H$ is the $i$th projection, the formalization proves
\begin{align}
    I_i^\dagger=P_i.
\end{align}
Block-diagonal and general block operators can then be defined by composing inclusions, entries, and projections:
\begin{lstlisting}[language=Lean]
noncomputable def blockDiagonal (A B : L ℋ) : L (HSum ℋ) :=
  hsumIncl ℋ 0 .comp (A.comp (hsumProj ℋ 0)) +
  hsumIncl ℋ 1 .comp (B.comp (hsumProj ℋ 1))

noncomputable def blockOp (A00 A01 A10 A11 : L ℋ) : L (HSum ℋ) :=
  hsumIncl ℋ 0 .comp (A00.comp (hsumProj ℋ 0)) +
  hsumIncl ℋ 0 .comp (A01.comp (hsumProj ℋ 1)) +
  hsumIncl ℋ 1 .comp (A10.comp (hsumProj ℋ 0)) +
  hsumIncl ℋ 1 .comp (A11.comp (hsumProj ℋ 1))
\end{lstlisting}
These definitions are the coordinate-free counterparts of
\begin{align}
    \operatorname{diag}[A,B]
    =
    \begin{pmatrix}
        A & 0\\
        0 & B
    \end{pmatrix},
    \qquad
    \operatorname{block}[A_{00},A_{01},A_{10},A_{11}]
    =
    \begin{pmatrix}
        A_{00} & A_{01}\\
        A_{10} & A_{11}
    \end{pmatrix}.
\end{align}

The basic positivity theorem for block-diagonal operators is:
\begin{lstlisting}[language=Lean]
theorem blockDiagonal_nonneg {A B : L ℋ} (hA : 0 ≤ A) (hB : 0 ≤ B) :
    0 ≤ blockDiagonal A B
\end{lstlisting}
The proof expands the quadratic form on a vector $(v_0,v_1)$:
\begin{align}
    \left\langle
        (v_0,v_1),
        \operatorname{diag}[A,B]
    \right\rangle
    =
    \langle v_0,Av_0\rangle+\langle v_1,Bv_1\rangle.
\end{align}
The formal proof has to account for the inner-product convention of \texttt{PiLp}, the adjointness of inclusions and projections, and the equivalence between operator positivity and nonnegativity of quadratic forms.

This block infrastructure is used in two ways. In the core of the proof of L\"owner--Heinz theorem, the analogous Schur-complement reasoning is carried out in $2\times2$ matrix algebras over a general $C^\ast$-algebra. In the Hilbert-space Jensen proof, explicit direct sums and block operators support unitary dilations and column-operator arguments. The mathematical idea is classical, but formalization forces a design decision: the same informal block notation must be split into a general $C^\ast$-algebra matrix interface for core results and a Hilbert-space direct-sum interface for dilation arguments.

\subsection{Hilbert--Schmidt Operator Space}
\label{subsec:hs}

The proof of the Lieb--Ando trace inequalities uses the standard idea of viewing $L(\mathcal H)$ itself as a Hilbert space. Informally, this is the Hilbert--Schmidt space with inner product
\begin{align}
    \langle X,Y\rangle_{\mathrm{HS}}
    =
    \Tr[X^\dagger Y].
\end{align}
In a matrix proof, this identification is often used without comment. In Lean, it must be turned into explicit structure: the operator space must be given a Hilbert-space structure, the trace formula for its inner product must be proved, and left and right multiplication must be represented as continuous linear operators on this Hilbert space.

The code introduces a type synonym for the Hilbert--Schmidt operator space:
\begin{lstlisting}[language=Lean]
def HSOp (ℋ : Type u) [NormedAddCommGroup ℋ] [InnerProductSpace ℂ ℋ] :
    Type u :=
  L ℋ
\end{lstlisting}
Mathematically, \texttt{HSOp H} has the same underlying elements as $L(\mathcal H)$, but it is used as a Hilbert space rather than as an algebra of operators acting on $\mathcal H$. The trace formula is packaged as:
\begin{lstlisting}[language=Lean]
lemma hsInner_eq_trace (X Y : L ℋ) :
    inner ℂ (ofOp X) (ofOp Y) =
      LinearMap.trace ℂ ℋ ((star X * Y).toLinearMap)
\end{lstlisting}
This is the bridge between Hilbert-space calculations on \texttt{HSOp H} and trace expressions on $L(\mathcal H)$.

Left and right multiplication are then defined as continuous linear maps:
\begin{lstlisting}[language=Lean]
noncomputable def leftMulHS (A : L ℋ) : HSOp ℋ →L[ℂ] HSOp ℋ :=
  LinearMap.toContinuousLinearMap
    { toFun := fun T => ofOp (A * toOp T)
      map_add' := fun T S => mul_add A (toOp T) (toOp S)
      map_smul' := fun z T => mul_smul_comm z A (toOp T) }

noncomputable def rightMulHS (B : L ℋ) : HSOp ℋ →L[ℂ] HSOp ℋ :=
  LinearMap.toContinuousLinearMap
    { toFun := fun T => ofOp (toOp T * B)
      map_add' := fun T S => add_mul (toOp T) (toOp S) B
      map_smul' := fun z T => smul_mul_assoc z (toOp T) B }
\end{lstlisting}
They represent
\begin{align}
    L_A(X)=AX,
    \qquad
    R_B(X)=XB.
\end{align}
They commute:
\begin{lstlisting}[language=Lean]
lemma leftMulHS_rightMulHS_commute (A B : L ℋ) :
    Commute (leftMulHS A) (rightMulHS B)
\end{lstlisting}
This is the formal version of associativity:
\begin{align}
    A(XB)=(AX)B.
\end{align}

The positivity and order behavior of these multiplication maps are also formalized:
\begin{lstlisting}[language=Lean]
lemma leftMulHS_nonneg {A : L ℋ} (hA0 : 0 ≤ A) :
    0 ≤ leftMulHS A

lemma rightMulHS_nonneg {B : L ℋ} (hB0 : 0 ≤ B) :
    0 ≤ rightMulHS B

lemma leftMulHS_le_leftMulHS {A B : L ℋ} (hAB : A ≤ B) :
    leftMulHS A ≤ leftMulHS B

lemma rightMulHS_le_rightMulHS {A B : L ℋ} (hAB : A ≤ B) :
    rightMulHS A ≤ rightMulHS B
\end{lstlisting}
For example, positivity of $L_A$ follows from
\begin{align}
    \langle X,L_A X\rangle_{\mathrm{HS}}
    =
    \Tr[X^\dagger A X],
\end{align}
and the right-hand side is nonnegative when $A\ge0$. Lean requires this reasoning to be decomposed into trace-inner-product conversion, positivity under conjugation, and nonnegativity of the trace of a positive operator.

The central compatibility lemma is functoriality of the continuous functional calculus for left and right multiplication:
\begin{lstlisting}[language=Lean]
lemma leftMulHS_cfcR
    (f : ℝ → ℝ) (A : L ℋ) (hA : IsSelfAdjoint A)
    (hf : ContinuousOn f (spectrum ℝ A)) :
    leftMulHS (cfcR f A) =
      cfcR f (leftMulHS A)

lemma rightMulHS_cfcR
    (f : ℝ → ℝ) (A : L ℋ) (hA : IsSelfAdjoint A)
    (hf : ContinuousOn f (spectrum ℝ A)) :
    rightMulHS (cfcR f A) =
      cfcR f (rightMulHS A)
\end{lstlisting}
Mathematically,
\begin{align}
    f(L_A)=L_{f(A)},
    \qquad
    f(R_A)=R_{f(A)}.
\end{align}
These identities are usually implicit in Hilbert--Schmidt proofs of trace inequalities. In Lean, they are substantive lemmas: the maps $A\mapsto L_A$ and $A\mapsto R_A$ must be shown to respect the algebraic and star structures in the form required by the CFC functoriality API.

With these lemmas, the trace expression
\begin{align}
    \Tr[A^sK^\dagger B^{1-s}K]
\end{align}
can be rewritten as a Hilbert--Schmidt quadratic form involving $L_A^sR_B^{1-s}$. This is the step that converts Lieb--Ando trace inequalities into operator-mean inequalities. The contribution of this subsection is therefore not a new analytic inequality, but the formal construction of the operator-space interface that makes the standard Hilbert--Schmidt method usable in Lean.

\subsection{Jensen's Operator Inequality}
\label{subsec:jensen}

The next layer is Jensen's operator inequality in the form needed for generalized perspectives. The mathematical input is an operator convex function $f$ satisfying $f(0)\le0$. The desired output is a finite Jensen inequality for contractions. If $A_j$ are self-adjoint with spectra in the domain of $f$, and if $X_j$ satisfy
\begin{align}
    \sum_j X_j^\dagger X_j \le I,
\end{align}
then
\begin{align}
    f\left[\sum_j X_j^\dagger A_jX_j\right]
    \le
    \sum_j X_j^\dagger f[A_j]X_j.
    \label{eq:jensen-finite}
\end{align}

The code separates the proof into named conditions. The first is operator convexity together with normalization:
\begin{lstlisting}[language=Lean]
def CondI (f : ℝ → ℝ) : Prop :=
  OperatorConvex f ∧ f 0 ≤ 0
\end{lstlisting}
The contraction form is:
\begin{lstlisting}[language=Lean]
def CondIV (f : ℝ → ℝ) : Prop :=
  ∀ {A X : L ℋ},
    IsSelfAdjoint A →
    spectrum ℝ A ⊆ Set.Ici (0 : ℝ) →
    ‖X‖ ≤ 1 →
    cfcR f (star X * A * X) ≤ star X * cfcR f A * X
\end{lstlisting}
and the two-operator form is:
\begin{lstlisting}[language=Lean]
def CondV (f : ℝ → ℝ) : Prop :=
  ∀ {A B X Y : L ℋ},
    IsSelfAdjoint A → IsSelfAdjoint B →
    spectrum ℝ A ⊆ Set.Ici (0 : ℝ) →
    spectrum ℝ B ⊆ Set.Ici (0 : ℝ) →
    star X * X + star Y * Y ≤ 1 →
    cfcR f (star X * A * X + star Y * B * Y)
      ≤ star X * cfcR f A * X + star Y * cfcR f B * Y
\end{lstlisting}
The main theorem packages the implication from the uniform operator convexity condition to the two-operator Jensen form:
\begin{lstlisting}[language=Lean]
theorem theorem_2_5_2_i_all_imp_v {f : ℝ → ℝ}
    (hf : CondIAll f) :
    CondV f
\end{lstlisting}
The use of \texttt{CondIAll} is a design choice. The proof moves between the original Hilbert space, a direct sum, and other Hilbert-space constructions. A scalar function property stated only for one fixed operator algebra would not be sufficient. The uniform statement allows the same convexity hypothesis to be applied after changing the ambient space.

The proof follows the Hansen--Pedersen dilation argument. First, a contraction $X$ is dilated to a unitary operator on $\mathcal H\oplus\mathcal H$ using the defect operators
\begin{align}
    (I-X^\dagger X)^{1/2},
    \qquad
    (I-XX^\dagger)^{1/2}.
\end{align}
The exact block arrangement is chosen to match the code. The proof then applies operator convexity to a conjugate of a block-diagonal operator, extracts the upper-left block, and uses $f(0)\le0$ to remove the defect term. This gives the contraction form.

Second, the two-operator form is obtained by packaging $X$ and $Y$ into a column operator
\begin{align}
    Wv=(Xv,Yv).
\end{align}
The hypothesis $X^\dagger X+Y^\dagger Y\le I$ says exactly that $W$ is a contraction. Applying the contraction form to the block-diagonal operator $\operatorname{diag}[A,B]$ gives Eq.~\eqref{eq:jensen-finite} after expanding the block expressions and using compatibility of the continuous functional calculus with block-diagonal operators.

This proof route differs from some matrix presentations, where Jensen-type inequalities may be derived through conditional expectations, pinching, or finite-dimensional subalgebras~\cite{Carlen2010}. The dilation route is convenient for the Lean development because it reduces the proof to explicit block operators, contractions, CFC square roots, unitary conjugation, and block extraction. The implementation effort is concentrated in proving that compressed operators have the required self-adjointness and spectral-domain properties, that the column operator is a contraction, and that block extraction preserves the relevant order inequality.

\subsection{Generalized Perspective Functions}
\label{subsec:perspective}

The generalized perspective theorem is the key intermediate result between Jensen's inequality and the Lieb--Ando trace inequalities. Rather than proceeding directly from matrix trace inequalities to the DPI, the formalization proves a reusable two-variable operator theorem and then specializes it later.

The formalization first introduces predicates for joint convexity and joint concavity on restricted domains. A representative definition is:
\begin{lstlisting}[language=Lean]
def JointlyConvexOn (s : Set E) (t : Set F) (Φ : E → F → G) : Prop :=
  ∀ {A₁ A₂ : E} {B₁ B₂ : F} {θ : ℝ},
    A₁ ∈ s → A₂ ∈ s → B₁ ∈ t → B₂ ∈ t →
    0 ≤ θ → θ ≤ 1 →
    Φ ((1 - θ) • A₁ + θ • A₂)
        ((1 - θ) • B₁ + θ • B₂)
      ≤ (1 - θ) • Φ A₁ B₁ + θ • Φ A₂ B₂
\end{lstlisting}
The domains are explicit because the two variables have different requirements. The first variable may only be positive semidefinite, whereas the second variable must be strictly positive when inverse square roots are used.

The generalized perspective of scalar functions $f$ and $h$ is defined by the real continuous functional calculus:
\begin{lstlisting}[language=Lean]
noncomputable def hSqrt (h : ℝ → ℝ) (B : L ℋ) : L ℋ :=
  cfcR (fun x : ℝ ↦ (h x) ^ ((1 : ℝ) / 2)) B

noncomputable def hInvSqrt (h : ℝ → ℝ) (B : L ℋ) : L ℋ :=
  cfcR (fun x : ℝ ↦ (h x) ^ ((-1 : ℝ) / 2)) B

noncomputable def GeneralizedPerspective
    (f h : ℝ → ℝ) (A B : L ℋ) : L ℋ :=
  hSqrt h B * cfcR f (hInvSqrt h B * A * hInvSqrt h B) * hSqrt h B
\end{lstlisting}
Mathematically,
\begin{align}
    (f\Delta h)(A,B)
    =
    h(B)^{1/2}
    f\left[h(B)^{-1/2}Ah(B)^{-1/2}\right]
    h(B)^{1/2}.
    \label{eq:generalized-perspective}
\end{align}
The use of the continuous functional calculus is essential. No diagonalization or fixed basis is used in the definition.

The positive semidefinite and positive definite domains are represented as spectral-domain predicates:
\begin{lstlisting}[language=Lean]
def psdSet : Set (L ℋ) :=
  {A | IsSelfAdjoint A ∧ spectrum ℝ A ⊆ Set.Ici (0 : ℝ)}

def pdSet : Set (L ℋ) :=
  {A | IsSelfAdjoint A ∧ spectrum ℝ A ⊆ Set.Ioi (0 : ℝ)}
\end{lstlisting}
The strict positivity of the second variable is not merely technical. It ensures that $h(B)^{-1/2}$ is an ordinary CFC expression on a positive spectrum.

The main convexity theorem is:
\begin{lstlisting}[language=Lean]
theorem theorem_2_5_forward_jointlyConvexOn_psd_pd
    {f h : ℝ → ℝ}
    (hf : CondIAll f)
    (hconc : OperatorConcaveOn (Set.Ioi (0 : ℝ)) h)
    (hcont : ContinuousOn h (Set.Ioi (0 : ℝ)))
    (hpos : ∀ x ∈ Set.Ioi (0 : ℝ), 0 < h x) :
    JointlyConvexOn psdSet pdSet
      (fun A B => GeneralizedPerspective f h A B)
\end{lstlisting}
Thus, if $f$ is operator convex with $f(0)\le0$, and $h$ is operator concave, continuous, and strictly positive on $(0,\infty)$, then $(f\Delta h)(A,B)$ is jointly convex on the positive-semidefinite and positive-definite domains.

The proof follows the argument of Ref.~\cite{NIKOUFAR2013531}. Given
\begin{align}
    A=(1-\theta)A_1+\theta A_2,
    \qquad
    B=(1-\theta)B_1+\theta B_2,
\end{align}
set $H_j=h(B_j)$ and $H=h(B)$. Operator concavity of $h$ gives
\begin{align}
    H\ge (1-\theta)H_1+\theta H_2.
\end{align}
The proof constructs contraction weights
\begin{align}
    X_1=\sqrt{1-\theta},H_1^{1/2}H^{-1/2},
    \qquad
    X_2=\sqrt{\theta},H_2^{1/2}H^{-1/2},
\end{align}
and verifies
\begin{align}
    X_1^\dagger X_1+X_2^\dagger X_2\le I.
\end{align}
The two-operator Jensen inequality is then applied to these weights. After conjugating by $H^{1/2}$ and simplifying the CFC expressions, one obtains the joint convexity inequality for the generalized perspective.

The corresponding concavity theorem is obtained by applying the convexity theorem to $-f$:
\begin{lstlisting}[language=Lean]
theorem theorem_2_6_forward_jointlyConcaveOn_psd_pd
    {f h : ℝ → ℝ}
    (hfconc : OperatorConcaveAll f)
    (hf0 : 0 ≤ f 0)
    (hconc : OperatorConcaveOn (Set.Ioi (0 : ℝ)) h)
    (hcont : ContinuousOn h (Set.Ioi (0 : ℝ)))
    (hpos : ∀ x ∈ Set.Ioi (0 : ℝ), 0 < h x) :
    JointlyConcaveOn psdSet pdSet
      (fun A B => GeneralizedPerspective f h A B)
\end{lstlisting}

The contribution of this subsection is the reusable interface. In a natural-language proof without using Lean, one can often suppress the construction of the contraction weights and the domain checks for inverse square roots. In Lean, these are the main proof obligations. By isolating them in the generalized perspective theorem, later proofs can use joint convexity and concavity without reopening the Jensen and inverse-square-root bookkeeping.

\subsection{Operator Power Means}
\label{subsec:mean}

The generalized perspective theorem is next specialized to power functions. This produces the operator power means used in the Hilbert--Schmidt proof of the Lieb--Ando trace inequalities:
\begin{lstlisting}[language=Lean]
noncomputable def operatorPowerMean (α β : ℝ) (A B : L ℋ) : L ℋ :=
  GeneralizedPerspective
    (fun x : ℝ ↦ x ^ α)
    (fun x : ℝ ↦ x ^ β)
    A B
\end{lstlisting}
For positive definite $A$ and $B$, this is
\begin{align}
    M_{\alpha,\beta}(A,B)
    =
    B^{\beta/2}
    \left(
        B^{-\beta/2}AB^{-\beta/2}
    \right)^\alpha
    B^{\beta/2}.
    \label{eq:operator-power-mean}
\end{align}
All powers are interpreted by the real continuous functional calculus. The additional parameter $\beta$ is useful because the later trace inequalities involve powers of both variables.

The main formal statements are:
\begin{lstlisting}[language=Lean]
theorem operatorPowerMean_jointlyConcaveOn_pdSet
    {α β : ℝ}
    (halpha : α ∈ Set.Icc (0 : ℝ) 1)
    (hbeta : β ∈ Set.Icc (0 : ℝ) 1) :
    JointlyConcaveOn pdSet pdSet
      (operatorPowerMean α β)

theorem operatorPowerMean_jointlyConvexOn_pdSet
    {α β : ℝ}
    (halpha : α ∈ Set.Icc (1 : ℝ) 2)
    (hbeta : β ∈ Set.Icc (0 : ℝ) 1) :
    JointlyConvexOn pdSet pdSet
      (operatorPowerMean α β)
\end{lstlisting}
Thus $M_{\alpha,\beta}$ is jointly concave on the positive definite cone for $0\le\alpha\le1$ and $0\le\beta\le1$, and jointly convex for $1\le\alpha\le2$ and $0\le\beta\le1$.

The proof is an instantiation of the generalized perspective theorem. For concavity, one takes $f(x)=x^\alpha$ and $h(x)=x^\beta$, and uses the operator concavity of both functions in the range $[0,1]$. For convexity, one uses operator convexity of $x^\alpha$ in the range $[1,2]$ and operator concavity of $x^\beta$ in the range $[0,1]$. The scalar-side obligations, such as continuity and strict positivity of $x^\beta$ on $(0,\infty)$, are discharged separately.

This is a proof-engineering layer. The generalized perspective theorem contains the hard Jensen and inverse-square-root argument; the power-mean theorem packages the specialization needed later. In Sec.~\ref{subsec:lieb}, the operator $L_A^sR_B^{1-s}$ on the Hilbert--Schmidt space is treated as an instance of such a power mean for the commuting positive operators $L_A$ and $R_B$. This converts a trace inequality into an operator-mean inequality.

The use of \texttt{pdSet} is deliberate. Since Eq.~\eqref{eq:operator-power-mean} contains inverse powers of $B$, the formal theorem is stated on the strictly positive cone. Extending corresponding statements to positive semidefinite operators would require a separate limiting argument, which is not mixed into the core analytic proof.

\subsection{Lieb--Ando Trace Inequalities}
\label{subsec:lieb}

The final layer is the formalization of Lieb--Ando type trace inequalities. These are the analytic results needed to prove joint convexity and concavity of the sandwiched quasi-entropy. In the present development, they are derived from the preceding perspective and power-mean infrastructure, rather than by directly formalizing each endpoint trace inequality as a separate matrix-trace argument.

The real-valued trace functionals are represented by taking the real part of the complex trace:
\begin{lstlisting}[language=Lean]
noncomputable def traceRe (T : L ℋ) : ℝ :=
  Complex.re (LinearMap.trace ℂ ℋ T.toLinearMap)

noncomputable def liebTraceMap (s : ℝ) (K : L ℋ) (A B : L ℋ) : ℝ :=
  traceRe (A ^ s * star K * B ^ (1 - s) * K)

noncomputable def andoTraceMap (q r : ℝ) (K : L ℋ) (A B : L ℋ) : ℝ :=
  traceRe (A ^ q * star K * B ^ (-r) * K)
\end{lstlisting}
The explicit real part is necessary because the trace over a complex Hilbert space is complex-valued in \textit{Mathlib}, while convexity and concavity are order-theoretic statements over $\mathbb R$. For the positive expressions used in applications, the trace is real, but this fact must be proved rather than inferred by coercion.

The formalized Lieb-type statements are:
\begin{lstlisting}[language=Lean]
theorem liebTrace_jointlyConcaveOn_pdSet
    {s : ℝ} (hs0 : 0 < s) (hs1 : s < 1) (K : L ℋ) :
    JointlyConcaveOn pdSet pdSet (liebTraceMap s K)

theorem liebTrace_jointlyConvexOn_pdSet
    {s : ℝ} (hs1 : 1 ≤ s) (hs2 : s ≤ 2) (K : L ℋ) :
    JointlyConvexOn pdSet pdSet (liebTraceMap s K)
\end{lstlisting}
They state that
\begin{align}
    (A,B)\mapsto
    \Re\Tr[A^sK^\dagger B^{1-s}K]
\end{align}
is jointly concave for $0<s<1$ and jointly convex for $1\le s\le2$ on the positive definite cone.

The Ando-type convexity statement is:
\begin{lstlisting}[language=Lean]
theorem andoTrace_jointlyConvexOn_pdSet
    {q r : ℝ} (hq1 : 1 ≤ q) (hq2 : q ≤ 2)
    (hr0 : 0 ≤ r) (hr1 : r ≤ 1) (hqr : 1 ≤ q - r)
    (K : L ℋ) :
    JointlyConvexOn pdSet pdSet (andoTraceMap q r K)
\end{lstlisting}
Mathematically, this asserts joint convexity of
\begin{align}
    (A,B)\mapsto
    \Re\Tr[A^qK^\dagger B^{-r}K]
\end{align}
under the parameter conditions
\begin{align}
    1\le q\le2,\qquad
    0\le r\le1,\qquad
    1\le q-r.
\end{align}
The positive definite domain is essential because the expression contains the negative power $B^{-r}$.

The proof of Lieb's concavity theorem uses the Hilbert--Schmidt space. Let $L_A$ and $R_B$ be left and right multiplication:
\begin{align}
    L_A(X)=AX,\qquad R_B(X)=XB.
\end{align}
If $A,B\ge0$, then $L_A$ and $R_B$ are positive operators on the Hilbert--Schmidt space, and they commute. The CFC-compatibility lemmas give
\begin{align}
    f(L_A)=L_{f(A)},\qquad f(R_B)=R_{f(B)}.
\end{align}
Consequently, the trace expression can be rewritten as a Hilbert--Schmidt quadratic form:
\begin{align}
    \Re\Tr[A^sK^\dagger B^{1-s}K]
    =
    \Re\left\langle
        K,
        L_A^sR_B^{1-s}K
    \right\rangle_{\mathrm{HS}},
    \label{eq:lieb-hs-rewrite}
\end{align}
up to the convention for the order of multiplication. The operator $L_A^sR_B^{1-s}$ is then controlled by the joint concavity or convexity of the appropriate operator power mean. Applying the positive real functional
\begin{align}
    T\mapsto \Re\langle K,TK\rangle_{\mathrm{HS}}
\end{align}
turns the operator inequality into the desired trace inequality.

The proof of the Ando trace inequality reduces the expression with $B^{-r}$ to a Lieb-type convex trace expression by a power substitution. For $q>1$, set
\begin{align}
    \beta=\frac{r}{q-1}.
\end{align}
The hypotheses imply $0\le\beta\le1$, and
\begin{align}
    (B^\beta)^{1-q}=B^{-r}.
\end{align}
The Ando functional can therefore be rewritten as a Lieb-type convex functional evaluated at $(A,B^\beta)$. The power map $B\mapsto B^\beta$ is controlled by the L\"owner--Heinz interface, and the required order behavior is supplied by the negative-power and convexity results from Sec.~\ref{subsec:lowner}. The boundary case $q=1$ is handled separately, where the hypotheses force $r=0$.

This final layer clarifies how the formal proof architecture differs from the way these inequalities are usually invoked in the literature. In Frank and Lieb's proof of the DPI, Lieb's concavity theorem and Ando's convexity theorem enter as previously established analytic results~\cite{FrankLieb2013}. Carlen's exposition develops related trace inequalities through finite-dimensional matrix methods, Hilbert--Schmidt inner products, and operator convexity arguments~\cite{Carlen2010}. Our Lean development formalizes the required Lieb--Ando inequalities through a more modular chain of intermediate results: L\"owner--Heinz inequalities for the continuous functional calculus, Jensen's operator inequality, generalized perspectives, operator power means, and the Hilbert--Schmidt left- and right-multiplication interface. Thus, the endpoint trace inequalities are not treated as isolated formal targets. They are obtained as consequences of reusable operator-theoretic components whose hypotheses, domain restrictions, and compatibility lemmas are all made explicit. This does not change the mathematical endpoint, but it gives a proof architecture better suited to machine checking and later reuse in other formalizations of noncommutative trace inequalities.

\section{Quantum Entropic Quantities}
\label{sec:entropy}

This section develops the entropy-specific part of the library and applies it to the proof of the DPI.
The preceding section supplies the noncommutative trace-inequality infrastructure, in particular the Lieb--Ando inequalities obtained through generalized perspectives, operator power means, and the Hilbert--Schmidt left/right multiplication interface. We apply this infrastructure to the sandwiched quasi-entropy
\begin{align}
    Q_{\alpha}(\rho\|\sigma)
    =
    \Tr\left[
        \left(
            \sigma^{\frac{1-\alpha}{2\alpha}}
            \rho
            \sigma^{\frac{1-\alpha}{2\alpha}}
        \right)^\alpha
    \right],
\end{align}
and to the sandwiched R\'enyi relative entropy
\begin{align}
    D_{\alpha}(\rho\|\sigma)
    =
    \frac{1}{\alpha-1}
    \log\left[
        \frac{Q_{\alpha}(\rho\|\sigma)}{\Tr[\rho]}
    \right].
\end{align}
The analytic core is first formalized on the positive definite cone, where inverse powers, inverse square roots, generalized perspectives, and variational expressions can be handled directly through the real continuous functional calculus. The DPI for positive semidefinite operators is then formalized separately by introducing an extended-real-valued non-negative version of the divergence and proving its monotonicity under CPTP maps.

Regarding the technical contributions, the proof modifies and simplifies the strategy of Frank and Lieb~\cite{FrankLieb2013} in several points that are important for Lean formalization. First, the variational formulas for $Q_{\alpha}$ are proved using trace Young and reverse-Young inequalities, rather than by formalizing an Euler--Lagrange optimization argument on the positive cone as in Ref.~\cite{FrankLieb2013}. Second, the Lieb--Ando trace inequalities are supplied through the generalized-perspective and Hilbert--Schmidt operator-space infrastructure developed in the preceding section. Third, the positive definite analytic proof is separated from the support-sensitive positive semidefinite extension. This organization exposes the domain, positivity, invertibility, support, tensor-product, and measure-theoretic assumptions that are often implicit in informal proofs, and turns the proof into reusable Lean interfaces.

The section is organized as follows. In Sec.~\ref{subsec:young}, we formalize trace Young and reverse-Young inequalities and use them to obtain variational formulas for $Q_{\alpha}$. This corresponds to Lemma~4 of Frank and Lieb, but we use an inequality-based proof route rather than formalizing an Euler--Lagrange argument on the positive cone. In Sec.~\ref{subsec:tensor}, we prove tensor-product compatibility of real powers through star-algebra homomorphisms and the continuous functional calculus, which gives tensor multiplicativity of $Q_{\alpha}$ and additivity of $D_{\alpha}$. In Sec.~\ref{subsec:haar}, we formalize normalized Haar measure on finite-dimensional unitary groups, prove a twirling identity, and derive the Haar averaging formula used in the channel part of the proof. In Sec.~\ref{subsec:dpi}, we combine these components with joint convexity and concavity of $Q_{\alpha}$ to prove the DPI.

\subsection{Variational Formulas via Young and Reverse-Young Inequalities}
\label{subsec:young}

The first entropy-specific ingredient is a variational representation of $Q_{\alpha}$. Frank and Lieb prove the corresponding statement in Lemma~4 of Ref.~\cite{FrankLieb2013}. Their presentation includes an Euler--Lagrange characterization of the optimizer and also notes a Young-inequality-based route in the case $\alpha>1$. For the Lean formalization, we use the Young-inequality route as the main proof principle and additionally prove the reverse-Young counterpart needed for the regime $0<\alpha<1$, which is one of the technical contributions of this work to quantum information theory itself since it gives a simpler proof of the variational formulas for $Q_{\alpha}$ in the full parameter range used in the DPI. The argument avoids differentiability on the positive cone and first-order optimality conditions for noncommutative trace functionals. Instead, it reduces the proof to scalar inequalities, spectral decompositions, doubly stochastic overlap matrices, cyclicity of trace, and explicit equality witnesses.

The scalar input is a weighted Young inequality summed against a doubly stochastic matrix:
\begin{lstlisting}[language=Lean]
lemma weighted_young_finset {ι κ : Type*}
    [Fintype ι] [Fintype κ]
    (p q : ℝ) (hpq : p.HolderConjugate q)
    (x : ι → ℝ) (y : κ → ℝ) (c : ι → κ → ℝ)
    (hx : ∀ i, 0 ≤ x i)
    (hy : ∀ j, 0 ≤ y j)
    (hc : ∀ i j, 0 ≤ c i j)
    (hrow : ∀ i, (∑ j, c i j) = 1)
    (hcol : ∀ j, (∑ i, c i j) = 1) :
    (∑ i, ∑ j, x i * y j * c i j)
      ≤ (∑ i, x i ^ p) / p + (∑ j, y j ^ q) / q
\end{lstlisting}
Mathematically, this is the scalar Young inequality
\begin{align}
    xy
    \le
    \frac{x^p}{p}
    +
    \frac{y^q}{q},
    \qquad
    \frac{1}{p}+\frac{1}{q}=1,
\end{align}
summed against nonnegative coefficients $c_{ij}$ whose row sums and column sums are one. The doubly stochastic form is the one naturally produced by comparing two spectral decompositions. If
\begin{align}
    X=\sum_i x_i |e_i\rangle\langle e_i|,
    \qquad
    Y=\sum_j y_j |f_j\rangle\langle f_j|,
\end{align}
then $c_{ij}=|\langle e_i,f_j\rangle|^2$ is doubly stochastic and
\begin{align}
    \Tr[XY]
    =
    \sum_{i,j}x_i y_j c_{ij}.
\end{align}
This yields the trace Young inequality:
\begin{lstlisting}[language=Lean]
theorem trace_young_inequality {ℋ : Type u} [Qudit ℋ]
    {p q : ℝ} (hpq : p.HolderConjugate q)
    (X Y : L ℋ) (hX : X.IsPositive) (hY : Y.IsPositive) :
    Tr (X ∘ₗ Y) ≤ Tr (CFC.rpow X p) / p + Tr (CFC.rpow Y q) / q
\end{lstlisting}
In mathematical notation,
\begin{align}
    \Tr[XY]
    \le
    \frac{\Tr[X^p]}{p}
    +
    \frac{\Tr[Y^q]}{q}.
    \label{eq:trace-young}
\end{align}
The proof is elementary in finite dimension, but it is not a one-line simplification in Lean. The spectral decompositions, eigenvalue functions, overlap coefficients, row and column normalization, and the conversion between coordinate sums and the coordinate-free trace must all be represented explicitly.

The regime $0<\alpha<1$ requires a reverse-Young inequality. The corresponding formal statement is:
\begin{lstlisting}[language=Lean]
theorem trace_reverse_young_inequality {ℋ : Type u} [Qudit ℋ]
    {r s : ℝ}
    (hr : r < 0) (hs0 : 0 < s) (hs1 : s < 1)
    (hrs : 1 / r + 1 / s = 1)
    (M N : L ℋ)
    (hM : M.IsPositive) (hM_unit : IsUnit M)
    (hN : N.IsPositive) :
    Tr (CFC.rpow M r) / r + Tr (CFC.rpow N s) / s
      ≤ Tr (M ∘ₗ N)
\end{lstlisting}
The scalar core is
\begin{align}
    \frac{a^r}{r}
    +
    \frac{b^s}{s}
    \le
    ab,
    \qquad
    r<0,\quad 0<s<1,\quad \frac{1}{r}+\frac{1}{s}=1.
    \label{eq:reverse-young}
\end{align}
The hypothesis that $M$ is a unit records the invertibility needed for the negative power $M^r$. This is one of the points where the positive definite restriction is not merely convenient but structurally necessary for the chosen analytic proof route.

The sandwiched quasi-entropy is then defined through the real continuous functional calculus:
\begin{lstlisting}[language=Lean]
noncomputable def sandwichedQuasi (α : ℝ) (ρ σ : L ℋ) : ℂ :=
  Tr (CFC.rpow
    (CFC.rpow σ ((1 - α) / (2 * α)) * ρ *
     CFC.rpow σ ((1 - α) / (2 * α)))
    α)
\end{lstlisting}
This is the Lean counterpart of
\begin{align}
    Q_{\alpha}(\rho\|\sigma)
    =
    \Tr\left[
        \left(
            \sigma^{\frac{1-\alpha}{2\alpha}}
            \rho
            \sigma^{\frac{1-\alpha}{2\alpha}}
        \right)^\alpha
    \right].
\end{align}
Since the trace over a complex Hilbert space is complex-valued in the library, the formalization also records the real-valuedness of the relevant expressions under positivity hypotheses and uses real parts when formulating convexity or concavity statements.

The variational functional has the form:
\begin{lstlisting}[language=Lean]
noncomputable def quasiVar
    (α : ℝ) (ρ σ H : L ℋ) : ℂ :=
  (α : ℂ) * Tr (H * ρ)
    - ((α - 1 : ℝ) : ℂ) *
      Tr (CFC.rpow
        (CFC.rpow σ ((α - 1) / (2 * α)) * H *
         CFC.rpow σ ((α - 1) / (2 * α)))
        (α / (α - 1)))
\end{lstlisting}
For $\alpha>1$, the corresponding mathematical statement is
\begin{align}
    Q_{\alpha}(\rho\|\sigma)
    =
    \sup_{H>0}
    \left\{
        \alpha\Tr[H\rho]
        -
        (\alpha-1)
        \Tr\left[
            \left(
                \sigma^{\frac{\alpha-1}{2\alpha}}
                H
                \sigma^{\frac{\alpha-1}{2\alpha}}
            \right)^{\frac{\alpha}{\alpha-1}}
        \right]
    \right\}.
    \label{eq:lemma4-alpha-gt-one}
\end{align}
For $0<\alpha<1$, the same expression gives the corresponding infimum formula, with the direction of the bound determined by the sign of $\alpha-1$. In the formalization, the proof is split into universal inequalities valid for every admissible auxiliary operator $H$ and equality lemmas for the chosen optimizer. The optimizer is selected so that equality holds in the scalar Young or reverse-Young inequality after the noncommutative substitutions have been made.

This route differs from the main presentation of Lemma~4 in Frank and Lieb~\cite{FrankLieb2013}. Formalizing an Euler--Lagrange proof would require differentiability of trace functionals on the positive definite cone, characterization of critical points, and additional domain management for inverse powers. The Young-inequality route avoids these layers. It replaces variational calculus by order inequalities with explicit equality cases, making the proof more local and better aligned with the finite-dimensional spectral infrastructure already needed elsewhere in the development.

\subsection{Continuous Functional Calculus on Tensor Products}
\label{subsec:tensor}

The DPI proof uses tensor multiplicativity of the sandwiched quasi-entropy and tensor additivity of the sandwiched R\'enyi relative entropy. The basic operator identity is
\begin{align}
    (A\otimes B)^p
    =
    A^p\otimes B^p
    \label{eq:rpow-tensor-product}
\end{align}
for positive operators $A\in L(\mathcal H_1)$ and $B\in L(\mathcal H_2)$, with powers interpreted by the real continuous functional calculus. On paper this is often treated as immediate from simultaneous spectral reasoning for $A\otimes I$ and $I\otimes B$. In Lean, it must be proved as a theorem with explicit positivity, commutation, and functional-calculus compatibility hypotheses.

The proof avoids choosing eigenbases. Instead, it factors a tensor-product operator through one-sided tensoring maps. These maps are implemented as star-algebra homomorphisms:
\begin{lstlisting}[language=Lean]
noncomputable def rTensorStarAlgHom :
    L ℋ₁ →⋆ₐ[ℂ] L (ℋ₁ ⊗[ℂ] ℋ₂) where
  toFun f := f.rTensor ℋ₂
  map_one' := by
    ext x y
    rfl
  map_mul' f g := by
    ext x y
    simp
  map_star' f := by
    simp [LinearMap.star_eq_adjoint, LinearMap.adjoint_rTensor]

noncomputable def lTensorStarAlgHom :
    L ℋ₂ →⋆ₐ[ℂ] L (ℋ₁ ⊗[ℂ] ℋ₂) where
  toFun g := g.lTensor ℋ₁
  map_one' := by
    ext x y
    rfl
  map_mul' f g := by
    ext x y
    simp
  map_star' g := by
    simp [LinearMap.star_eq_adjoint, LinearMap.adjoint_lTensor]
\end{lstlisting}
These maps send $A$ to $A\otimes I$ and $B$ to $I\otimes B$, up to the tensor-factor conventions fixed by the code. The tensor product itself is then factored as
\begin{lstlisting}[language=Lean]
lemma map_eq_rTensor_mul_lTensor (f : L ℋ₁) (g : L ℋ₂) :
    (TensorProduct.map f g : L (ℋ₁ ⊗[ℂ] ℋ₂)) =
      rTensorStarAlgHom f * lTensorStarAlgHom g

lemma commute_rTensor_lTensor (f : L ℋ₁) (g : L ℋ₂) :
    Commute (rTensorStarAlgHom f) (lTensorStarAlgHom g)
\end{lstlisting}
Mathematically,
\begin{align}
    A\otimes B
    =
    (A\otimes I)(I\otimes B),
    \qquad
    (A\otimes I)(I\otimes B)
    =
    (I\otimes B)(A\otimes I).
\end{align}
Functoriality of the continuous functional calculus gives one-sided compatibility:
\begin{lstlisting}[language=Lean]
lemma rpow_rTensor (A : L ℋ₁) (p : ℝ) (hA : 0 ≤ A) :
    CFC.rpow (rTensorStarAlgHom A) p =
      rTensorStarAlgHom (CFC.rpow A p)

lemma rpow_lTensor (B : L ℋ₂) (p : ℝ) (hB : 0 ≤ B) :
    CFC.rpow (lTensorStarAlgHom B) p =
      lTensorStarAlgHom (CFC.rpow B p)
\end{lstlisting}
Combining these one-sided identities with commutation of the two tensor factors yields:
\begin{lstlisting}[language=Lean]
theorem rpow_tensorProduct (A : L ℋ₁) (B : L ℋ₂) (p : ℝ)
    (hA : 0 ≤ A) (hB : 0 ≤ B) :
    CFC.rpow (TensorProduct.map A B : L (ℋ₁ ⊗[ℂ] ℋ₂)) p =
      TensorProduct.map (CFC.rpow A p) (CFC.rpow B p)
\end{lstlisting}
This theorem is then used twice in the proof of tensor multiplicativity of $Q_{\alpha}$. For positive operators $\rho_1,\sigma_1\in L(\mathcal H_1)$ and $\rho_2,\sigma_2\in L(\mathcal H_2)$,
\begin{align}
    Q_{\alpha}(\rho_1\otimes\rho_2\|\sigma_1\otimes\sigma_2)
    =
    Q_{\alpha}(\rho_1\|\sigma_1)
    Q_{\alpha}(\rho_2\|\sigma_2).
    \label{eq:qalpha-tensor}
\end{align}
The corresponding logarithmic statement is additivity:
\begin{align}
    D_{\alpha}(\rho_1\otimes\rho_2\|\sigma_1\otimes\sigma_2)
    =
    D_{\alpha}(\rho_1\|\sigma_1)
    +
    D_{\alpha}(\rho_2\|\sigma_2).
    \label{eq:dalpha-tensor}
\end{align}
In the DPI proof, this additivity removes the maximally mixed ancillary state introduced by Haar averaging, because
\begin{align}
    Q_{\alpha}\left(
        \frac{I_E}{d_E}
        \middle\|
        \frac{I_E}{d_E}
    \right)
    =
    1,
    \qquad
    D_{\alpha}\left(
        \frac{I_E}{d_E}
        \middle\|
        \frac{I_E}{d_E}
    \right)
    =
    0.
\end{align}

The formal contribution of this subsection is not the tensor identity itself, which is standard, but the CFC-level interface proving it without introducing tensor-product eigenbases. The star-algebra-homomorphism proof is closer to the abstractions used by \textit{Mathlib}: it uses functoriality of the continuous functional calculus, one-sided tensoring maps, and commutation of the corresponding subalgebras. As a result, the theorem can be reused in later formalizations requiring powers of tensor-product positive operators, not only in the proof of DPI for the sandwiched R\'enyi relative entropy.

\subsection{Haar Measure and Unitary Averaging}
\label{subsec:haar}

The next structural ingredient is the Haar-averaging identity that converts a partial trace into an average over unitary conjugations. In Frank and Lieb's proof~\cite{FrankLieb2013}, this is a standard finite-dimensional averaging step. In Lean, it requires several explicit components: a compact topological group structure on the unitary group, normalized Haar measure as a probability measure, Bochner integration of operator-valued functions, and a twirling identity as an equality in an operator space.

The normalized Haar measure on the finite-dimensional unitary group is represented as follows:
\begin{lstlisting}[language=Lean]
noncomputable def haarUnitary (ℋ : Type u) [Qudit ℋ] [Nontrivial ℋ] :
    Measure (unitary (L ℋ)) :=
  haarMeasure (unitaryPC ℋ)

instance haarUnitary_isHaarMeasure [Nontrivial ℋ] :
    IsHaarMeasure (haarUnitary ℋ) := by
  unfold haarUnitary
  infer_instance

instance haarUnitary_isProbabilityMeasure [Nontrivial ℋ] :
    IsProbabilityMeasure (haarUnitary ℋ) :=
  ⟨by change haarMeasure (unitaryPC ℋ) Set.univ = 1
      have : haarMeasure (unitaryPC ℋ) (unitaryPC ℋ : Set _) = 1 := haarMeasure_self
      simpa [unitaryPC] using this⟩
\end{lstlisting}
Here \texttt{unitary (L H)} is the group of unitary elements of the operator algebra $L(\mathcal H)$, and \texttt{unitaryPC H} packages it with the compact topological group structure required by the Haar-measure API. The probability-measure instance records the normalization of Haar measure. This normalization is essential because the later use of Jensen's inequality requires a convex average.

The central averaging result is the twirling formula:
\begin{lstlisting}[language=Lean]
theorem twirl_eq_smul_one (X : L ℋ) :
    integral (fun u : unitary (L ℋ) =>
      (u : L ℋ) * X * star (u : L ℋ)) (haarUnitary ℋ)
    =
    (((Module.finrank ℂ ℋ : ℂ) ^ (-1)) * LinearMap.trace ℂ ℋ X) • (1 : L ℋ)
\end{lstlisting}
Mathematically,
\begin{align}
    \int_{\mathcal U(\mathcal H)}
        U X U^\dagger dU
    =
    \frac{\Tr[X]}{\dim[\mathcal H]} I.
    \label{eq:twirl}
\end{align}
The proof uses invariance of Haar measure to show that the average commutes with every unitary, a finite-dimensional Schur-type argument to identify the average as a scalar multiple of the identity, and trace preservation under unitary conjugation to determine the scalar. The Lean proof must additionally establish measurability and integrability of the operator-valued integrand and justify the exchange of the trace with the Bochner integral.

The formalization also provides Jensen inequalities for Bochner integrals of jointly convex and jointly concave functions:
\begin{lstlisting}[language=Lean]
theorem jointly_convex_integral_le
    {α : Type*} [MeasurableSpace α] {μ : Measure α} [IsProbabilityMeasure μ]
    {S T : Set (L ℋ)} (hS : Convex ℝ S) (hT : Convex ℝ T)
    (hSc : IsClosed S) (hTc : IsClosed T)
    {f : L ℋ → L ℋ → ℝ}
    (hf_conv : JointlyConvexOn S T f)
    (hf_cont : ContinuousOn (Function.uncurry f) (S ×ˢ T))
    {g₁ g₂ : α → L ℋ}
    (hg₁ : ∀ᵐ x ∂μ, g₁ x ∈ S)
    (hg₂ : ∀ᵐ x ∂μ, g₂ x ∈ T)
    (hg₁_int : Integrable g₁ μ) (hg₂_int : Integrable g₂ μ)
    (hfg_int : Integrable (fun x => f (g₁ x) (g₂ x)) μ)
    (_hmem₁ : ∫ x, g₁ x ∂μ ∈ S) (_hmem₂ : ∫ x, g₂ x ∂μ ∈ T) :
    f (∫ x, g₁ x ∂μ) (∫ x, g₂ x ∂μ) ≤
      ∫ x, f (g₁ x) (g₂ x) ∂μ

theorem jointly_concave_le_integral
    {α : Type*} [MeasurableSpace α] {μ : Measure α} [IsProbabilityMeasure μ]
    {S T : Set (L ℋ)} (hS : Convex ℝ S) (hT : Convex ℝ T)
    (hSc : IsClosed S) (hTc : IsClosed T)
    {f : L ℋ → L ℋ → ℝ}
    (hf_conc : JointlyConcaveOn S T f)
    (hf_cont : ContinuousOn (Function.uncurry f) (S ×ˢ T))
    {g₁ g₂ : α → L ℋ}
    (hg₁ : ∀ᵐ x ∂μ, g₁ x ∈ S)
    (hg₂ : ∀ᵐ x ∂μ, g₂ x ∈ T)
    (hg₁_int : Integrable g₁ μ) (hg₂_int : Integrable g₂ μ)
    (hfg_int : Integrable (fun x => f (g₁ x) (g₂ x)) μ)
    (_hmem₁ : ∫ x, g₁ x ∂μ ∈ S) (_hmem₂ : ∫ x, g₂ x ∂μ ∈ T) :
    (∫ x, f (g₁ x) (g₂ x) ∂μ) ≤
      f (∫ x, g₁ x ∂μ) (∫ x, g₂ x ∂μ)
\end{lstlisting}
These are the integral analogues of finite Jensen inequalities. The omitted hypotheses include measurability, integrability, probability-measure normalization, and domain preservation of the averaged operators. These conditions are routine in a finite-dimensional hand proof but must be explicit in a theorem prover because Bochner integration is formulated in a general measure-theoretic setting.

Combining the twirling formula with the partial trace yields the Stinespring--Haar averaging identity used in the DPI proof. A representative statement is:

\begin{lstlisting}[language=Lean]
theorem stinespring_haar_eq
    (U : unitary (L (ℋ₁ ⊗[ℂ] ℋ₂))) (τ : L ℋ₁) (γ : L ℋ₂) :
    TensorProduct.map
      ((Module.finrank ℂ ℋ₁ : ℂ)⁻¹ • LinearMap.id (M := ℋ₁))
      (QuantumChannel.Tr₂
        ((U : L (ℋ₁ ⊗[ℂ] ℋ₂)) *
         TensorProduct.map τ γ *
         star (U : L (ℋ₁ ⊗[ℂ] ℋ₂)))) =
    ∫ u : unitary (L ℋ₁),
      TensorProduct.map ((u : L ℋ₁)) (LinearMap.id (M := ℋ₂)) *
        ((U : L (ℋ₁ ⊗[ℂ] ℋ₂)) *
         TensorProduct.map τ γ *
         star (U : L (ℋ₁ ⊗[ℂ] ℋ₂))) *
      TensorProduct.map (star (u : L ℋ₁)) (LinearMap.id (M := ℋ₂))
    ∂(haarUnitary ℋ₁)
\end{lstlisting}
In mathematical notation, this is the identity
\begin{align}
    \frac{I_E}{d_E}\otimes \Tr_E[X]
    =
    \int_{\mathcal U(E)}
        (u\otimes I)X(u^\dagger\otimes I)du,
    \label{eq:stinespring-haar-basic}
\end{align}
applied to a Stinespring-dilated input. Thus, if $\Phi$ is represented by a Stinespring dilation, then
\begin{align}
    \frac{I_E}{d_E}\otimes \Phi[\gamma]
    =
    \int_{\mathcal U(E)}
        (u\otimes I)
        V\gamma V^\dagger
        (u^\dagger\otimes I)du,
    \label{eq:stinespring-haar-channel}
\end{align}
with tensor-factor conventions determined by the formal partial trace.

Compared with the natural-language proof, the averaging idea is unchanged. The formal contribution is that the normalized Haar measure, the twirling identity, the partial trace, and Jensen's inequality for Bochner integrals are all separated into reusable lemmas. This makes the averaging step a general interface for channel monotonicity arguments rather than a proof-specific manipulation inside the DPI argument.

\subsection{Data Processing Inequality for the Sandwiched R\'enyi Relative Entropy}
\label{subsec:dpi}

The primary theorem of this section is the DPI for the sandwiched R\'enyi relative entropy, including its positive semidefinite formulation. The formalized analytic core is first stated for positive definite operators.
This restriction is a proof-engineering choice aligned with the continuous-functional-calculus representation of powers: all inverse powers and inverse square roots are ordinary operators on the strictly positive cone.
The support-conditioned statement for positive semidefinite operators is then formalized through an extended-real-valued non-negative divergence, with the regularization argument corresponding to replacing $\sigma$ by $\sigma+\varepsilon I$ and taking $\varepsilon\downarrow0$.

The theorem can be stated mathematically as follows.
\begin{theorem}[Data processing inequality on the positive definite cone]
Let $\Phi:L(\mathcal H_1)\to L(\mathcal H_2)$ be completely positive and trace preserving. Let $\rho$ and $\sigma$ be density operators on $\mathcal H_1$, and assume that $\sigma$ is positive definite. For $\alpha\in[1/2,1)\cup(1,\infty)$,
\begin{align}
    D_{\alpha}(\Phi[\rho]\|\Phi[\sigma])
    \le
    D_{\alpha}(\rho\|\sigma).
\end{align}
\end{theorem}
A representative Lean-level statement has the form:
\begin{lstlisting}[language=Lean]
theorem sandwichedRenyiDiv_monotone
    {ℋ : Type u} [Qudit ℋ] [Nontrivial ℋ]
    {�� : Type u} [Qudit ��] [Nontrivial ��]
    (E : CPTP ℋ ��) {α : ℝ}
    (hα_ge : (1 : ℝ) / 2 ≤ α) (hα_ne1 : α ≠ 1)
    {ρ σ : L ℋ}
    (hρ : ρ ∈ pdSetLM (ℋ := ℋ)) (hσ : σ ∈ pdSetLM (ℋ := ℋ))
    (hEρ : E.toFun ρ ∈ pdSetLM (ℋ := ��))
    (hEσ : E.toFun σ ∈ pdSetLM (ℋ := ��)) :
    sandwichedRenyiDiv α (E.toFun ρ) (E.toFun σ) ≤
      sandwichedRenyiDiv α ρ σ
\end{lstlisting}
The statement follows the predicate-based style used throughout the development: the operators are ambient elements of $L(\mathcal H)$, and membership in the strictly positive domain is a hypothesis.
The displayed theorem is the positive-definite analytic core. The development also formalizes the perturbed non-negative version and the final extended-real-valued DPI for positive semidefinite operators.

The proof is assembled from the preceding formal components. First, the variational formulas of Sec.~\ref{subsec:young} reduce joint convexity and concavity of $Q_{\alpha}$ to Lieb--Ando trace inequalities. In schematic form:
\begin{lstlisting}[language=Lean]
theorem sandwichedQuasi_re_jointlyConvex {α : ℝ} (hα : 1 < α) :
    JointlyConvexOn (pdSetLM (ℋ := ℋ)) (pdSetLM (ℋ := ℋ))
      (fun ρ σ => (sandwichedQuasi (ℋ := ℋ) α ρ σ).re)

theorem sandwichedQuasi_re_jointlyConcave
    {α : ℝ} (hα_ge : 1 / 2 ≤ α) (hα_lt : α < 1) :
    JointlyConcaveOn (pdSetLM (ℋ := ℋ)) (pdSetLM (ℋ := ℋ))
      (fun ρ σ => (sandwichedQuasi (ℋ := ℋ) α ρ σ).re)
\end{lstlisting}
This is the formal counterpart of Proposition~3 in Frank and Lieb~\cite{FrankLieb2013}: $Q_{\alpha}$ is jointly concave for $1/2\le\alpha<1$ and jointly convex for $\alpha>1$. In the Lean proof, the parameter transformations connecting the variational formulas to the Lieb--Ando trace inequalities are explicit.

Second, the channel is replaced by a Stinespring dilation and the partial trace is rewritten using Haar averaging. The isometric part is handled separately through invariance of the quasi-entropy under isometric conjugation:
\begin{align}
    Q_{\alpha}(V\rho V^\dagger\|V\sigma V^\dagger)
    =
    Q_{\alpha}(\rho\|\sigma).
    \label{eq:dpi-isometry-invariance}
\end{align}
The partial trace part is handled by Eq.~\eqref{eq:stinespring-haar-channel}. This separation is useful formally because isometry invariance is an algebraic and continuous-functional-calculus argument, whereas Haar averaging is a measure-theoretic argument involving Bochner integrals.

Third, Jensen's inequality is applied to the Haar average. For $\alpha>1$, joint convexity gives
\begin{align}
    Q_{\alpha}\left(
        \frac{I_E}{d_E}\otimes\Phi[\rho]
        \middle|
        \frac{I_E}{d_E}\otimes\Phi[\sigma]
    \right)
    \le
    Q_{\alpha}(V\rho V^\dagger\|V\sigma V^\dagger)
    =
    Q_{\alpha}(\rho\|\sigma).
    \label{eq:dpi-intermediate-alpha-gt-one}
\end{align}
For $1/2\le\alpha<1$, joint concavity gives the reverse inequality at the quasi-entropy level:
\begin{align}
    Q_{\alpha}\left(
        \frac{I_E}{d_E}\otimes\Phi[\rho]
        \middle|
        \frac{I_E}{d_E}\otimes\Phi[\sigma]
    \right)
    \ge
    Q_{\alpha}(\rho\|\sigma).
    \label{eq:dpi-intermediate-alpha-lt-one}
\end{align}
The reversal is necessary because $(\alpha-1)^{-1}<0$ in this regime.

Fourth, tensor multiplicativity removes the ancillary maximally mixed state. At the quasi-entropy level, the formal theorem has the form:
\begin{lstlisting}[language=Lean]
theorem sandwichedQuasi_tensor
    (α : ℝ) (ρ₁ σ₁ : L ℋ₁) (ρ₂ σ₂ : L ℋ₂)
    (hρ₁ : 0 ≤ ρ₁) (hσ₁ : 0 ≤ σ₁)
    (hρ₂ : 0 ≤ ρ₂) (hσ₂ : 0 ≤ σ₂) :
    sandwichedQuasi α (TensorProduct.map ρ₁ ρ₂ : L (ℋ₁ ⊗[ℂ] ℋ₂))
        (TensorProduct.map σ₁ σ₂) =
      sandwichedQuasi α ρ₁ σ₁ * sandwichedQuasi α ρ₂ σ₂
\end{lstlisting}
The corresponding divergence statement is:
\begin{lstlisting}[language=Lean]
theorem sandwichedRenyiDiv_tensor
    (α : ℝ) (ρ₁ σ₁ : L ℋ₁) (ρ₂ σ₂ : L ℋ₂)
    (hρ₁ : 0 ≤ ρ₁) (hσ₁ : 0 ≤ σ₁) (hρ₂ : 0 ≤ ρ₂) (hσ₂ : 0 ≤ σ₂)
    (hQ₁ : (sandwichedQuasi α ρ₁ σ₁).re ≠ 0)
    (hQ₂ : (sandwichedQuasi α ρ₂ σ₂).re ≠ 0)
    (hT₁ : (Tr ρ₁).re ≠ 0) (hT₂ : (Tr ρ₂).re ≠ 0)
    (hQ₁im : (sandwichedQuasi α ρ₁ σ₁).im = 0)
    (hQ₂im : (sandwichedQuasi α ρ₂ σ₂).im = 0)
    (hT₁im : (Tr ρ₁).im = 0) (hT₂im : (Tr ρ₂).im = 0) :
    sandwichedRenyiDiv α (TensorProduct.map ρ₁ ρ₂ : L (ℋ₁ ⊗[ℂ] ℋ₂))
        (TensorProduct.map σ₁ σ₂) =
      sandwichedRenyiDiv α ρ₁ σ₁ + sandwichedRenyiDiv α ρ₂ σ₂
\end{lstlisting}
Since the maximally mixed ancillary pair contributes zero divergence, the tensor factor introduced by Haar averaging disappears.

Finally, the proof passes from the quasi-entropy inequality to the logarithmic divergence. This step uses monotonicity of the real logarithm, positivity of the quasi-entropy, and the sign of $(\alpha-1)^{-1}$. Lean requires each of these facts explicitly: the argument of $\log$ must be positive, the complex trace must be related to the intended real expression, and multiplying by $(\alpha-1)^{-1}$ must preserve or reverse the inequality according to the parameter regime.

The resulting proof has the same mathematical endpoint as the Frank--Lieb proof but a different formal proof structure. Lemma~4 is proved through trace Young and reverse-Young inequalities rather than by formalizing Euler--Lagrange equations. Proposition~3 is obtained from the Lieb--Ando inequalities developed through generalized perspectives and Hilbert--Schmidt left/right multiplication. Tensor multiplicativity is proved through star-algebra homomorphisms and the continuous functional calculus rather than by an implicit spectral calculation. Haar averaging is expressed through normalized Haar measure, a twirling theorem, and Jensen inequalities for Bochner integrals. These choices make the formal proof longer than the natural-language proof, but they expose its logical dependencies as reusable components for later machine-checked results in quantum entropy theory.

\section{Conclusion}
\label{sec:conclusion}

We have presented a Lean 4 library for the theory of quantum information. The library is not a one-off encoding of a single theorem, but a reusable formal infrastructure for finite-dimensional quantum systems, operator theory, and quantum entropic analysis. As a central demonstration of the library, we formalized the data processing inequality (DPI) for the sandwiched R\'enyi relative entropy for positive semidefinite operators on finite-dimensional quantum systems.
On the quantum-information side, the library provides a coordinate-free interface for finite-dimensional systems, states, channels, tensor products, partial traces, vectorization, Choi operators, Kraus representations, and Stinespring representations. On the mathematical side, it formalizes a hierarchy of operator-theoretic tools needed for noncommutative trace inequalities, including operator monotonicity and convexity, block-operator arguments, Hilbert--Schmidt operator spaces, Jensen's operator inequality, generalized perspectives, operator power means, Lieb--Ando trace inequalities, trace Young and reverse-Young inequalities, tensor-product compatibility of the continuous functional calculus, Haar integration, and unitary averaging. These components are not specific to the sandwiched R\'enyi relative entropy. They provide reusable theorem-prover infrastructure for future formalizations of quantum entropic inequalities and related results in quantum information theory.

The development also makes a technical contribution to quantum information theory itself. Rather than translating an existing natural-language proof into Lean line by line, we reorganized the proof of the DPI into stable intermediate interfaces with explicit hypotheses. For example, the part of the proof in Ref.~\cite{FrankLieb2013} concerning the variational formulas for $Q_{\alpha}$ is reformulated using Young and reverse-Young inequalities. This avoids formalizing the Euler--Lagrange optimization argument on the positive cone used in Ref.~\cite{FrankLieb2013}, and instead reduces the proof to scalar inequalities, spectral decompositions, and explicit equality witnesses. We also use generalized perspective functions~\cite{NIKOUFAR2013531} to give a direct formal route to the Lieb--Ando trace inequalities. The Hilbert--Schmidt operator-space construction then converts the required trace functionals into quadratic forms of operator means, allowing the trace inequalities to be derived from operator-mean inequalities.
A further structural point is the separation between the positive definite analytic core and the positive semidefinite final theorem. The core proof of the DPI is first formalized on the positive definite cone, where inverse powers, inverse square roots, and generalized perspectives can be handled directly through the real continuous functional calculus. This isolates the main noncommutative analytic argument from support projections and boundary conventions, and makes explicit the domain, positivity, invertibility, and support assumptions that are often left implicit in informal arguments. The positive semidefinite statement is then formalized by introducing an extended-real-valued non-negative version of the divergence and proving its monotonicity under CPTP maps. This organization makes the Lean library precise, modular, and compatible with the standard \textit{Mathlib} interfaces.

More broadly, this work illustrates how quantum information theory can be developed as machine-checkable mathematical infrastructure. In the AI era, a textbook for quantum information theory should not merely be a collection of human-readable explanations and informal proofs. It should also be accompanied by formal libraries that expose definitions, assumptions, and logical dependencies in a form that proof assistants can inspect and verify. Such libraries can serve as machine-checkable reference texts for both human researchers and AI-assisted tools. Lean provides a substrate for this purpose. Just as a compiler checks that a program is well typed, Lean's kernel checks that each accepted proof term has the claimed theorem as its type. We expect that such verified infrastructure will become increasingly important as quantum information theory, formal proof assistants, and AI-assisted mathematical research become more closely integrated.

\begin{acknowledgments}
    HY acknowledges Alex Meiburg for discussions.
    HY was supported by JST CREST Grant Number JPMJCR25I5, JST [Moonshot R\&D] [Grant Number JPMJMS256J], JST PRESTO Grant Number JPMJPR23FC, and Faculty Research Funding from Google Quantum AI\@. SS was supported by JST CREST JPMJCR25I5, JST BOOST JPMJBY24E2, and JSPS KAKENHI 24K21316.
\end{acknowledgments}

\bibliography{citation}

\end{document}